\documentclass[twocolumn, aps, prd, superscriptaddress,floatfix]{revtex4-1}

\usepackage{multirow}
\usepackage{amsmath,amssymb}
\usepackage{graphicx,bm}
\usepackage{slashed}
\usepackage{epstopdf}
\usepackage{ulem} 
\usepackage[usenames]{color}
\usepackage{float}
\usepackage{braket}
\usepackage{appendix}
\usepackage{hyperref}

\renewcommand\sout{\bgroup \color{red} \ULdepth=-.5ex \ULset}

\begin{document}

\title{Diquarkyonic matter: quarks, diquarks and baryons}

\author{Aaron Park}
\email{aaron.park@yonsei.ac.kr}\affiliation{Department of Physics and Institute of Physics and Applied Physics, Yonsei University, Seoul 03722, Korea}
\author{Su Houng~Lee}\email{suhoung@yonsei.ac.kr}\affiliation{Department of Physics and Institute of Physics and Applied Physics, Yonsei University, Seoul 03722, Korea}
\date{\today}
\begin{abstract}
In this work, we investigate the color-spin interaction of a quark, a diquark and a baryon with their surrounding baryons and/or quark matter. We extend our previous work \cite{Park:2022qew} by increasing the maximum number of surrounding baryons to 5 and additionally consider all possible diquark probes that are immersed in such surroundings.
This is accomplished by classifying all possible flavor and spin states of the resulting multiquark configuration in both the flavor SU(2) and SU(3) symmetric cases. We also discuss the three-body confinement potential and show that this does not contribute to the outcome. Furthermore, we find that a quark becomes more stable than a baryon when the number of surrounding baryons is three or more. Finally, when we consider the internal color-spin factor of a probe, our results show that the effects of the color-spin interaction of a multiquark configuration is consistent with the so-called diquarkyonic configuration.

\end{abstract}

\maketitle

\section{Introduction}

Recent neutron star observations \cite{Demorest:2010bx, Antoniadis:2013pzd, NANOGrav:2019jur}, which constrain the equation of state(EOS) of matter, lead us to understand the properties of dense matter in a new perspective.  
Gravitaional wave astrophysics also has begun to provide EOS constraints measuring tidal deformability from the neutron star merger \cite{LIGOScientific:2017vwq, LIGOScientific:2018cki}. Additionally, recent analyses by NICER from pulsar J0740+6620 \cite{Miller:2021qha, Riley:2021pdl} estimated radii of neutron star as $R_{1.4}\simeq 12.45$ km, $R_{2.08}\simeq 12.35$ km, whose large mass and radius supports the stiff
evolution of EOS around the core density. The quarkyonic matter configuration was introduced to explain this stiff evolution of EOS \cite{McLerran:2007qj}.

The quarkyonic matter was originally discussed in the limit of large $N_C$ quantum chromodynamics(QCD) description of cold dense matter. The quarkyonic configuration with shell-like phase space distribution of baryon can be generated through the hard-core repulsive interaction between nucleons \cite{Jeong:2019lhv,Duarte:2020xsp,Duarte:2020kvi}. This new picture explains stiffening of EOS in a high density region before the transition   to quark matter.
The essential ingredient of this configuration is the hard-core repulsion and this can be well understood in terms of quark level interaction with Pauli blocking \cite{Park:2019bsz}.

In order to probe possible phases in a quark model point of view, we showed that the short distance repulsion between the quark and the baryon is smaller than that between two baryons in the lowest energy channel using a constituent quark model \cite{Park:2021hqb}. Also, the point where the behavior occurs is consistent with the quarkyonic configuration. When the relevant scales are such that the quark mass differences can be neglected so that all quark involved have similar distributions, the color-color interaction can be neglect. In this case, the most important factor determining the repulsive force at short distance is the color-spin interaction.  The color-spin interaction is a key factor explaining the repulsive core between 
nucleons \cite{Park:2019bsz} and also is an important ingredient in examining the bound state of exotic hadrons such as tetraquark \cite{Yoon:2022voo}. 

In Ref.\cite{Park:2021hqb}, by calculating the energy of a quark and a baryon using both the color-spin and color-color interaction,  it was shown that a quarkyoniclike phase can appear when the density is high. This state  differs from the existing quarkyonic configuration as 
we only considered the interaction energy of the $d$ quark with the surrounding baryons leaving out the interaction of the  $(ud)$ diquark that together with the $d$ quark comprised the initial neutron in matter.  This was so because such configuration  costs the least amount of excitation energy: hence  called a quarkyoniclike configuration to distinguish it from the quarkyonic configuration where all interaction of all three quarks inside the neutron is important for long range excitation mode in the momentum space.
As a result, if not only quarks but also diquarks can appear together, it can be called a so-called diquarkyonic matter, a new phase that is distinct from the existing configuration.
However, if the diquark can exist as an independent state, the interaction that the diquark experiences from the surrounding baryons needs to be calculated as well.

In this paper, we extend our previous work \cite{Park:2022qew} in several respects. First of all, the maximum number of surrounding baryons is increased from 3 to 5. Second, 
in addition to considering the interaction of the quark and the most attractive diquark, which we will henceforth call probes, with the surrounding, we will consider all additional diquark probes to investigate which one is the most stable diquark in dense matter. Third, when we discuss about color-color interaction, we add the investigation on three-body confinement potential. Fourth, a case for flavor decuplet baryon is added. Also, we describe in more detail how we determine the allowed states for each case.

In this work, we assume the spatial part of the multiquark wave function to be totally symmetric. Also, since the color state of entire multiquark state is determined from the probe, we can determine the possible flavor and spin state to satisfy the Pauli exclusion principle.

This paper is organized as follows. In Sec.\ref{Color-spin interaction}, the formulas for the color-spin interaction factor are introduced for both flavor SU(2) and SU(3). In Sec.\ref{color-color interaction}, we discuss three-body confinement potentials induced by three-gluon exchange. In Sec.\ref{multibaryon}, we construct the multibaryon states. These states are used for both correlated surrounding baryons and the multiquark states when a probe is a baryon. In Sec.\ref{multiquark}, we construct the multiquark states when a probe is a quark, a diquark, three correlated diquarks and a baryon, respectively. In Sec.\ref{free-quark}, we consider the case where the surrounding is a free quark gas. In Sec.\ref{results}, we analyze the results using all the states we constructed. Sec.\ref{summary} is devoted to summary and concluding remarks.

\section{Color-spin interaction}
\label{Color-spin interaction}

In this work, we will assume that all quarks  occupy the same spatial configuration.  This means that the spatial potential will be universal for any pair so that the relative strength is determined only by the color and/or spin factors. This factor for the color-spin interaction is defined  as follows.
\begin{align}
V_{CS}&=-\sum_{i<j}^n \frac{1}{m_i m_j}\lambda^c_i \lambda^c_j \sigma_i \cdot \sigma_j \nonumber \\
&\equiv \frac{1}{m_u^2} H_{CS},
\label{color-spin}
\end{align}
where $\lambda^c_i$, $m_i$, $m_u$ are respectively the color SU(3) Gell-Mann matrices, the constituent quark mass of the $i$'th quark, and the constutient quark mass of $u,d$ quarks. For flavor SU(3) symmetric case, color-spin factor $H_{CS}$ can be easily calculated by the following formula.

\begin{align}
  H_{CS}&=-\sum_{i<j}^n \lambda^c_i \lambda^c_j  \sigma_i \cdot \sigma_j \nonumber \\
  &= n(n-10) + \frac{4}{3}S(S+1)+4C_F+2C_C, \nonumber\\
  4C_F &= \frac{4}{3}(p_1^2+p_2^2+3p_1+3p_2+p_1 p_2),
\label{CSF-1}
\end{align}
where $C_F$ is the first kind of the Casimir operator of the flavor SU(3) and $p_i$ is the number of columns containing $i$ boxes in a column in Young diagram. For flavor SU(2) case, Eq.(\ref{CSF-1}) reduces to the following formula.
\begin{align}
  H_{CS} = \frac{4}{3}n(n-6) + \frac{4}{3}S(S+1)+ 4I(I+1) + 2C_C,
\label{CSF-2}
\end{align}
where $I$ is the total isospin.

\section{Color-color interaction}
\label{color-color interaction}

We can also consider the color-color interaction, which is responsible for the confining and Coulomb type of interactions. However, as we showed in Ref.\cite{Park:2022qew}, this interaction between the quarks in the probe and those in the color singlet configuration cancel out. Additionally, we can also consider color interaction induced by three-gluon exchange \cite{Dmitrasinovic:2001nu, Pepin:2001is,Papp:2002xh}. There are two types of three-body color operators in color SU(3) which are $d^{abc}F^a_i F^b_j F^c_k$ and $f^{abc}F^a_i F^b_j F^c_k$. We can represent these operators using the permutations of symmetric group as follows \cite{Park:2015nha}.

\begin{align}
  d^{abc}F^a_i F^b_j F^c_k &= \frac{1}{4}[(ijk)+(ikj)] + \frac{1}{9} \label{d-type}\nonumber \\
  & \quad -\frac{1}{6}[(ij)+(ik)+(jk)], \\
  f^{abc}F^a_i F^b_j F^c_k &= \frac{i}{4}[(ijk)-(ikj)], \label{f-type}
\end{align}
where $(ij)$ and $(ijk)$ are 2-cycle and 3-cycle permutations, respectively. 

Let's first consider $d$-type three-body confinement potential. Using the fact that $\sum_{i<j<k} d^{abc} F^a_i F^b_j F^c_k$ is Casimir operator, we can calculate its eigenvalue considering only the normal Young-Yamanouchi basis and the axial distance among $i,j$ and $k$. 

\begin{widetext}
\begin{align}
    \sum_{i<j<k} d^{abc}F_i^a F_j^b F_k^c
    &= \frac{p_1}{27} -\frac{p_1^2}{18} + \frac{p_1^3}{54} + \frac{13p_2}{54}-\frac{5p_1 p_2}{36} + \frac{p_1^2 p_2}{36} -\frac{2p_2^2}{9} -\frac{p_1 p_2^2}{36} -\frac{p_2^3}{54} + \frac{10p_3}{9},
\end{align}
\end{widetext}
where $p_i$ is the number of columns containing $i$ boxes in a column in Young diagram. As we can see, the above formula is linear in $p_3$, which is the number of singlet baryons in the corresponding color state. It shows that the $d$-type three-body confinement potentials between the singlet and the others cancel out. This holds not only for SU(3), but in general. We represent the eigenvalue of $d$-type three-body confinement potential for SU(4), SU(5) and SU(6) and how to calculate it in the appendix.

Now, let's consider the $f$-type three-body confinement potential. As we can see in Eq.(\ref{f-type}), it is not difficult to check that the diagonal component of it vanishes. However, since $f$-type three-body confinement potential is not Casimir operator, we need to check its off-diagonal element. By calculating it directly using the permutation matrices, we can check that all elements vanish only if the color state is $[m,m,\cdots,m]$ type for SU($N$), where $m$ is any integer and the number of rows are smaller or equal to $N$. However, in this work, since we focus on the diagonal component of interaction, we only consider the color-spin factor to study the interaction between a probe and surrounding baryons in dense matter.

\section{Multibaryon states}
\label{multibaryon}

In this section, we first construct the multibaryon states. To calculate the interaction which a probe experiences from the surrounding baryons, we subtract the internal color-spin factor of surrounding baryons from the entire multiquark state. When there are two or more baryons, there can be multiple possible states so it is necessary to consider all possible cases.

\subsection{Two baryons}
\label{2b}

The color state of two baryons should be color singlet. Hence, the remaining part of the wave function can be determined using the conjugate form of the color state.
The color and flavor $\otimes$ spin states of two baryons are as follows.\\
\\
Color : $\begin{tabular}{|c|c|}
  \hline
  \quad \quad & \quad \quad  \\
  \hline
  \quad \quad & \quad \quad \\
  \hline
  \quad \quad & \quad \quad \\
  \hline
\end{tabular}$, \quad
Flavor $\otimes$ Spin : $\begin{tabular}{|c|c|c|}
  \hline
  \quad \quad & \quad \quad & \quad \quad  \\
  \hline
  \quad \quad & \quad \quad & \quad \quad  \\
  \hline
\end{tabular}.$\\

The flavor $\otimes$ spin coupling state [3,3] with SU(6) can be decomposed into the states with the flavor SU(3) and the spin SU(2) as follows.
\begin{align}
[3,3]_{FS} =& [6]_F \otimes [3,3]_S + [5,1]_F \otimes [4,2]_S + [4,2]_F \otimes [5,1]_S \nonumber\\
&+ [4,2]_F \otimes [3,3]_S + [3,3]_F \otimes [6]_S  \nonumber\\
 &  + [3,3]_F \otimes [4,2]_S + [3,2,1]_F \otimes [5,1]_S \nonumber\\
 & + [3,2,1]_F \otimes [4,2]_S+ [2,2,2]_F \otimes [3,3]_S.
\end{align}

However, all flavor and spin states in the above are not possible as two baryons states. Since we only consider the flavor octet baryon in this work, the possible flavor states should be determined using the outer product as follows.

Flavor states of 2 baryons :
\begin{align}
\mathbf{8}\times \mathbf{8} = \mathbf{1}+\mathbf{8}_{(m=2)}+\mathbf{10}+\overline{\mathbf{10}}+\mathbf{27}.
\end{align}
Then the possible flavor and spin states of two baryons are listed as follows.
\begin{widetext}
Flavor and spin : $\begin{tabular}{|c|c|c}
  \cline{1-2}
  \quad \quad & \quad \quad  \\
  \cline{1-2}
  \quad \quad & \quad \quad \\
  \cline{1-2}
  \quad \quad & \quad \quad \\
  \cline{1-2}
  \multicolumn{3}{c}{$\mathbf{1}(S=0)$}
\end{tabular}$,
$\begin{tabular}{|c|c|c|c}
  \cline{1-3}
  \quad \quad & \quad \quad & \quad \quad \\
  \cline{1-3}
  \quad \quad & \quad \quad \\
  \cline{1-2}
  \quad \quad \\
  \cline{1-1}
  \multicolumn{4}{c}{$\mathbf{8}(S=1)$}
\end{tabular}$,
$\begin{tabular}{|c|c|c|c|c}
  \cline{1-4}
  \quad \quad & \quad \quad & \quad \quad & \quad \quad \\
  \cline{1-4}
  \quad \quad \\
  \cline{1-1}
  \quad \quad \\
  \cline{1-1}
  \multicolumn{5}{c}{$\mathbf{10}(S=1)$}
\end{tabular}$\quad ,
$\begin{tabular}{|c|c|c|c}
  \cline{1-3}
  \quad \quad & \quad \quad & \quad \quad \\
  \cline{1-3}
  \quad \quad & \quad \quad & \quad \quad \\
  \cline{1-3}
  \multicolumn{4}{c}{$\mathbf{\overline{10}}(S=1)$}
\end{tabular}$,
$\begin{tabular}{|c|c|c|c|c}
  \cline{1-4}
  \quad \quad & \quad \quad & \quad \quad & \quad \quad \\
  \cline{1-4}
  \quad \quad & \quad \quad \\
  \cline{1-2}
  \multicolumn{5}{c}{$\mathbf{27}(S=0)$}.
\end{tabular}$\\
\end{widetext}
Here, the Young diagrams are the flavor states and the possible spin states are shown in the parentheses.

\subsection{Three baryons}
\label{3b}

The color and flavor $\otimes$ spin states of three baryons are as follows.\\
\\
Color : $\begin{tabular}{|c|c|c|}
  \hline
  \quad \quad & \quad \quad & \quad \quad \\
  \hline
  \quad \quad & \quad \quad & \quad \quad \\
  \hline
  \quad \quad & \quad \quad & \quad \quad \\
  \hline
\end{tabular}$, \quad
Flavor $\otimes$ Spin : $\begin{tabular}{|c|c|c|}
  \hline
  \quad \quad & \quad \quad & \quad \quad  \\
  \hline
  \quad \quad & \quad \quad & \quad \quad  \\
  \hline
  \quad \quad & \quad \quad & \quad \quad  \\
  \hline
\end{tabular}.$\\

The flavor $\otimes$ spin coupling state [3,3,3] with SU(6) can be decomposed into the states with the flavor SU(3) and the spin SU(2) as follows.
\begin{widetext}
\begin{align}
[3,3,3]_{FS} =&[6,3]_F \otimes [6,3]_S + [6,2,1]_F \otimes [5,4]_S + [5,4]_F \otimes [5,4]_S + [5,3,1]_F \otimes [7,2]_S + [5,3,1]_F \otimes [6,3]_S \nonumber\\
 & + [5,3,1]_F \otimes [5,4]_S + [5,2,2]_F \otimes [6,3]_S + [4,4,1]_F \otimes [6,3]_S+ [4,3,2]_F \otimes [8,1]_S + [4,3,2]_F \otimes [7,2]_S \nonumber\\
 & + [4,3,2]_F \otimes [6,3]_S + [4,3,2]_F \otimes [5,4]_S + [3,3,3]_F \otimes [9]_S + [3,3,3]_F \otimes [7,2]_S + [3,3,3]_F \otimes [6,3]_S.
\end{align}
Similar to the two baryons case, the possible flavor states of three baryons are determined as follows.

Flavor states of 3 baryons :
\begin{align}
  \mathbf{8} \times \mathbf{8} \times \mathbf{8} =& \mathbf{1}_{(m=2)}+\mathbf{8}_{(m=8)}+\mathbf{10}_{(m=4)}+\mathbf{\overline{10}}_{(m=4)} +\mathbf{27}_{(m=6)}+\mathbf{35}_{(m=2)}+\mathbf{\overline{35}}_{(m=2)}+\mathbf{64}.
\end{align}

Flavor and spin :
$\begin{tabular}{|c|c|c|c}
  \cline{1-3}
  \quad \quad & \quad \quad & \quad \quad  \\
  \cline{1-3}
  \quad \quad & \quad \quad & \quad \quad \\
  \cline{1-3}
  \quad \quad & \quad \quad & \quad \quad \\
  \cline{1-3}
  \multicolumn{4}{c}{$\mathbf{1}(S=\frac{3}{2})$}
\end{tabular}$,
$\begin{tabular}{|c|c|c|c|}
  \cline{1-4}
  \quad \quad & \quad \quad & \quad \quad & \quad \quad  \\
  \cline{1-4}
  \quad \quad & \quad \quad & \quad \quad \\
  \cline{1-3}
  \quad \quad & \quad \quad \\
  \cline{1-2}
  \multicolumn{4}{l}{$\mathbf{8}(S=\frac{1}{2},\frac{3}{2})$}
\end{tabular}$,
$\begin{tabular}{|c|c|c|c|c|}
  \hline
  \quad \quad & \quad \quad & \quad \quad & \quad \quad & \quad \quad  \\
  \hline
  \quad \quad & \quad \quad  \\
  \cline{1-2}
  \quad \quad & \quad \quad \\
  \cline{1-2}
  \multicolumn{5}{c}{$\mathbf{10}(S=\frac{3}{2})$}
\end{tabular}$,
$\begin{tabular}{|c|c|c|c|}
  \hline
  \quad \quad & \quad \quad & \quad \quad & \quad \quad  \\
  \hline
  \quad \quad & \quad \quad & \quad \quad & \quad \quad \\
  \hline
  \quad \quad \\
  \cline{1-1}
  \multicolumn{4}{c}{$\mathbf{\bar{10}}(S=\frac{3}{2})$}
\end{tabular}$,
$\begin{tabular}{|c|c|c|c|c|}
  \hline
  \quad \quad & \quad \quad & \quad \quad & \quad \quad & \quad \quad  \\
  \hline
  \quad \quad & \quad \quad & \quad \quad \\
  \cline{1-3}
  \quad \quad \\
  \cline{1-1}
  \multicolumn{5}{c}{$\mathbf{27}(S=\frac{1}{2},\frac{3}{2})$}
\end{tabular}$,
$\begin{tabular}{|c|c|c|c|c|c|}
  \hline
  \quad \quad & \quad \quad & \quad \quad & \quad \quad & \quad \quad & \quad \quad \\
  \hline
  \quad \quad & \quad \quad \\
  \cline{1-2}
  \quad \quad \\
  \cline{1-1}
  \multicolumn{6}{c}{$\mathbf{35}(S=\frac{1}{2})$}
\end{tabular}$,
$\begin{tabular}{|c|c|c|c|c|}
  \hline
  \quad \quad & \quad \quad & \quad \quad & \quad \quad & \quad \quad  \\
  \hline
  \quad \quad & \quad \quad & \quad \quad & \quad \quad \\
  \cline{1-4}
  \multicolumn{5}{c}{$\mathbf{\bar{35}}(S=\frac{1}{2})$}
\end{tabular}$,
$\begin{tabular}{|c|c|c|c|c|c|}
  \hline
  \quad \quad & \quad \quad & \quad \quad & \quad \quad & \quad \quad & \quad \quad \\
  \hline
  \quad \quad & \quad \quad & \quad \quad \\
  \cline{1-3}
  \multicolumn{6}{c}{$\mathbf{64}(S=\frac{3}{2})$ } \\
\end{tabular} $. \\
\end{widetext}

\subsection{Four baryons}
\label{4b}

The color and flavor $\otimes$ spin states of four baryons are as follows.\\

Color : $\begin{tabular}{|c|c|c|c|}
  \cline{1-4}
  \quad \quad & \quad \quad & \quad \quad & \quad \quad \\
  \cline{1-4}
  \quad \quad & \quad \quad & \quad \quad & \quad \quad \\
  \cline{1-4}
  \quad \quad & \quad \quad & \quad \quad & \quad \quad \\
  \cline{1-4}
\end{tabular}$, \quad
Flavor $\otimes$ Spin : $\begin{tabular}{|c|c|c|}
  \cline{1-3}
  \quad \quad & \quad \quad & \quad \quad  \\
  \cline{1-3}
  \quad \quad & \quad \quad & \quad \quad  \\
  \cline{1-3}
  \quad \quad & \quad \quad & \quad \quad  \\
  \cline{1-3}
  \quad \quad & \quad \quad & \quad \quad  \\
  \cline{1-3}
\end{tabular}.$\\
The flavor $\otimes$ spin coupling state [3,3,3,3] with SU(6) can be decomposed into the states with the flavor SU(3) and the spin SU(2) as follows.
\begin{widetext}
\begin{align}
[3,3,3,3]_{FS} &=[6,6]_F \otimes [6,6]_S + [6,5,1]_F \otimes [7,5]_S + [6,4,2]_F \otimes [8,4]_S + [6,4,2]_F \otimes [6,6]_S + [6,3,3]_F \otimes [9,3]_S \nonumber\\
  &\quad + [6,3,3]_F \otimes [7,5]_S + [5,5,2]_F \otimes [7,5]_S + [5,4,3]_F \otimes [8,4]_S + [5,4,3]_F \otimes [7,5]_S + [4,4,4]_F \otimes [6,6]_S.
\end{align}
The flavor and spin states of four octet baryons are as follows.\\

Flavor states of 4 baryons :
\begin{align}
  \mathbf{8} \times \mathbf{8} \times \mathbf{8} \times \mathbf{8}
  =& \mathbf{1}_{(m=8)}+\mathbf{8}_{(m=32)}+\mathbf{10}_{(m=20)}+\mathbf{\overline{10}}_{(m=20)}+\mathbf{27}_{(m=33)}+\mathbf{28}_{(m=2)}+\mathbf{\overline{28}}_{(m=2)} +\mathbf{35}_{(m=15)} +\mathbf{\overline{35}}_{(m=15)} \nonumber\\
  & +\mathbf{64}_{(m=12)} +\mathbf{81}_{(m=3)}+ \mathbf{125}.
\end{align}
Flavor and spin :
$\begin{tabular}{|c|c|c|c|}
  \hline
  \quad \quad & \quad \quad & \quad \quad & \quad \quad \\
  \cline{1-4}
  \quad \quad & \quad \quad & \quad \quad & \quad \quad \\
  \cline{1-4}
  \quad \quad & \quad \quad & \quad \quad & \quad \quad \\
  \cline{1-4}
  \multicolumn{4}{c}{$\mathbf{1}(S=0)$}
\end{tabular}$\quad,
$\begin{tabular}{|c|c|c|c|c|}
  \hline
  \quad \quad & \quad \quad & \quad \quad & \quad \quad & \quad \quad \\
  \cline{1-5}
  \quad \quad & \quad \quad & \quad \quad & \quad \quad \\
  \cline{1-4}
  \quad \quad & \quad \quad & \quad \quad \\
  \cline{1-3}
  \multicolumn{5}{c}{$\mathbf{8}(S=1,2)$}
\end{tabular}$,
$\begin{tabular}{|c|c|c|c|c|c|}
  \hline
  \quad \quad & \quad \quad & \quad \quad & \quad \quad & \quad \quad & \quad \quad \\
  \hline
  \quad \quad & \quad \quad & \quad \quad \\
  \cline{1-3}
  \quad \quad & \quad \quad & \quad \quad \\
  \cline{1-3}
  \multicolumn{6}{c}{$\mathbf{10}(S=1)$}
\end{tabular}$,
$\begin{tabular}{|c|c|c|c|c|}
  \hline
  \quad \quad & \quad \quad & \quad \quad & \quad \quad & \quad \quad \\
  \hline
  \quad \quad & \quad \quad & \quad \quad & \quad \quad & \quad \quad \\
  \hline
  \quad \quad & \quad \quad \\
  \cline{1-2}
  \multicolumn{5}{c}{$\mathbf{\bar{10}}(S=1)$}
\end{tabular}$,
$\begin{tabular}{|c|c|c|c|c|c|}
  \hline
  \quad \quad & \quad \quad & \quad \quad & \quad \quad & \quad \quad & \quad \quad \\
  \hline
  \quad \quad & \quad \quad & \quad \quad & \quad \quad \\
  \cline{1-4}
  \quad \quad & \quad \quad \\
  \cline{1-2}
  \multicolumn{6}{c}{$\mathbf{27}(S=0,2)$}
\end{tabular}$,
$\begin{tabular}{|c|c|c|c|c|c|}
  \hline
  \quad \quad & \quad \quad & \quad \quad & \quad \quad & \quad \quad & \quad \quad \\
  \hline
  \quad \quad & \quad \quad & \quad \quad & \quad \quad & \quad \quad \\
  \cline{1-5}
  \quad \quad \\
  \cline{1-1}
  \multicolumn{6}{c}{$\mathbf{\bar{35}}(S=1)$}
\end{tabular}$,
$\begin{tabular}{|c|c|c|c|c|c|}
  \hline
  \quad \quad & \quad \quad & \quad \quad & \quad \quad & \quad \quad & \quad \quad \\
  \hline
  \quad \quad & \quad \quad & \quad \quad & \quad \quad & \quad \quad & \quad \quad \\
  \hline
  \multicolumn{6}{c}{$\mathbf{\overline{28}}(S=0)$ } \\
\end{tabular} $. \\
\end{widetext}

\subsection{Five baryons}
\label{5b}

The color and flavor $\otimes$ spin states of five baryons are as follows.\\

Color : $\begin{tabular}{|c|c|c|c|c|}
  \hline
  \quad \quad & \quad \quad & \quad \quad & \quad \quad & \quad \quad \\
  \hline
  \quad \quad & \quad \quad & \quad \quad & \quad \quad & \quad \quad \\
  \hline
  \quad \quad & \quad \quad & \quad \quad & \quad \quad & \quad \quad \\
  \hline
\end{tabular}$, \quad
Flavor $\otimes$ Spin : $\begin{tabular}{|c|c|c|}
  \hline
  \quad \quad & \quad \quad & \quad \quad  \\
  \hline
  \quad \quad & \quad \quad & \quad \quad  \\
  \hline
  \quad \quad & \quad \quad & \quad \quad  \\
  \hline
  \quad \quad & \quad \quad & \quad \quad  \\
  \hline
  \quad \quad & \quad \quad & \quad \quad  \\
  \hline
\end{tabular}.$\\

The flavor $\otimes$ spin coupling state [3,3,3,3,3] with SU(6) can be decomposed into the states with the flavor SU(3) and the spin SU(2) as follows.
\begin{align}
[3,3,3,3,3]_{FS} =[6,6,3]_F \otimes [9,6]_S + [6,5,4]_F \otimes [8,7]_S.
\end{align}

\begin{widetext}
Flavor states of 5 baryons :
\begin{align}
  \mathbf{8} \times \mathbf{8} \times \mathbf{8} \times \mathbf{8} \times \mathbf{8}
  =& \mathbf{1}_{(m=32)}+\mathbf{8}_{(m=145)}+\mathbf{10}_{(m=100)}+\mathbf{\overline{10}}_{(m=100)}+\mathbf{27}_{(m=180)}+\mathbf{28}_{(m=20)}+\mathbf{\overline{28}}_{(m=20)} +\mathbf{35}_{(m=100)} \nonumber\\
   &+\mathbf{\overline{35}}_{(m=100)} +\mathbf{64}_{(m=94)} +\mathbf{80}_{(m=5)} +\overline{\mathbf{80}}_{(m=5)} +\mathbf{81}_{(m=36)} +\overline{\mathbf{81}}_{(m=36)} +\mathbf{125}_{(m=20)} +\mathbf{154}_{(m=4)} \nonumber\\
   & +\overline{\mathbf{154}}_{(m=4)}+ \mathbf{216}.
\end{align}
The flavor and spin states of four octet baryons are as follows.\\

Flavor and spin :
$\begin{tabular}{|c|c|c|c|c|c|}
  \hline
  \quad \quad & \quad \quad & \quad \quad & \quad \quad & \quad \quad & \quad \quad \\
  \hline
  \quad \quad & \quad \quad & \quad \quad & \quad \quad & \quad \quad & \quad \quad \\
  \hline
  \quad \quad & \quad \quad & \quad \quad \\
  \cline{1-3}
  \multicolumn{5}{c}{$\mathbf{\bar{10}}(S=\frac{3}{2})$}
\end{tabular}$,
$\begin{tabular}{|c|c|c|c|c|c|}
  \hline
  \quad \quad & \quad \quad & \quad \quad & \quad \quad & \quad \quad & \quad \quad \\
  \hline
  \quad \quad & \quad \quad & \quad \quad & \quad \quad & \quad \quad \\
  \cline{1-5}
  \quad \quad & \quad \quad & \quad \quad & \quad \quad \\
  \cline{1-4}
  \multicolumn{6}{c}{$\mathbf{8}(S=\frac{1}{2})$}
\end{tabular}$.\\
\end{widetext}
Theoretically, we can construct six baryons states with totally symmetric spatial wave function. However, if we add a probe to the six baryons state, then the total state does not satisfy the Pauli principle. Therefore, in this study, we limit the number of surrounding baryons up to five.

\section{Flavor, color and spin states of a multiquark system}
\label{multiquark}

In this section, we construct the entire multiquark states which are composed of multibaryon and a probe. As a probe, we consider a quark, a baryon, a diquark and three correlated diquarks.

\subsection{Quark case}

\subsubsection{One baryon + one quark}
\label{1b1q}

Let us consider the multiquark system consisting of one baryon and one quark. In this work, we assume that the spatial part of a wave function is totally symmetric and the surrounding baryons are in flavor octet states. Since a baryon is a color singlet, the color state of four quarks should be a triplet.  Then, the flavor $\otimes$ spin coupling state of four quarks should be the conjugate of the color state to satisfy the Pauli exclusion principle. \\

Color : $\begin{tabular}{|c|c|}
  \cline{1-2}
  \quad \quad & \quad \quad   \\
  \cline{1-2}
  \quad \quad \\
  \cline{1-1}
  \quad \quad \\
  \cline{1-1}
\end{tabular}$, \quad
Flavor $\otimes$ Spin : $\begin{tabular}{|c|c|c|}
  \cline{1-3}
  \quad \quad & \quad \quad & \quad \quad  \\
  \cline{1-3}
  \quad \quad \\
  \cline{1-1}
\end{tabular}$.\\

The flavor $\otimes$ spin coupling state [3,1] with SU(6) can be decomposed into the states with the flavor SU(3) and the spin SU(2) as follows.
\begin{align}
[3,1]_{FS} =& [4]_F \otimes [3,1]_S + [3,1]_F \otimes [4]_S + [3,1]_F \otimes [3,1]_S \nonumber \\
& + [3,1]_F \otimes [2,2]_S + [2,2]_F \otimes [3,1]_S \nonumber \\
&+ [2,1,1]_F \otimes [3,1]_S + [2,1,1]_F \otimes [2,2]_S.
\label{PFS-1b1q}
\end{align}

Meanwhile, we can determine the possible flavor states of multiquark system as follows.\\
\\
Flavor states of 4 quarks :
\begin{align}
  \mathbf{8}\times \mathbf{3}=\mathbf{3}+\overline{\mathbf{6}}+\mathbf{15}.
\label{PF-1b1q}
\end{align}
Since only the above flavor states are allowed, we can classify all possible flavor and spin states of a four quarks configuration from the Eq.(\ref{PFS-1b1q}) and (\ref{PF-1b1q}) as follows.\\

Flavor and spin : $\begin{tabular}{|c|c|c}
  \cline{1-2}
  \quad \quad & \quad \quad  \\
  \cline{1-2}
  \quad \quad \\
  \cline{1-1}
  \quad \quad \\
  \cline{1-1}
  \multicolumn{3}{c}{$\mathbf{3}(S=0,1)$}
\end{tabular}$,
$\begin{tabular}{|c|c|c}
  \cline{1-2}
  \quad \quad & \quad \quad   \\
  \cline{1-2}
  \quad \quad & \quad \quad\\
  \cline{1-2}
  \multicolumn{3}{c}{$\mathbf{\overline{6}}(S=0,1)$}
\end{tabular}$,
$\begin{tabular}{|c|c|c|c}
  \cline{1-3}
  \quad \quad & \quad \quad & \quad \quad  \\
  \cline{1-3}
  \quad \quad \\
  \cline{1-1}
  \multicolumn{4}{c}{$\mathbf{15}(S=0,1)$}
\end{tabular}$.\\

Now, we investigate the relative magnitude of the interaction which a quark inside the probe sees from the surrounding $n$ baryons using the following formula.

\begin{align}
  \Delta H_{CS}^{n\mathrm{b}+p} &= H_{CS}^{n\mathrm{b}+p} - H_{CS}^{n\mathrm{b}} - H_{CS}^{p},  \label{2b-1} \\
  \Delta H_{CS}^{\mathrm{avg}} &= \frac{1}{n_p n \sum_{C,F,S} d_{CFS}}\sum_{C,F,S}d_{CFS}\Delta H_{CS}^{n\mathrm{b}+p}, \label{2b-2} \\
  d_{CFS} &= d_C d_F d_S m_{FS}.
\end{align}
Here, $n\mathrm{b}$ and $p$ in the superscripts represent $n$ external baryons and the probe, respectively.  The probe will be a baryon, a quark, a diquark or three correlated diquarks.  $n_p$ is the number of quarks in the probe.  We will investigate cases with $n=1,2,3,4,5$, and also consider the case where  $n\mathrm{b}$ is replaced by a single quark so as to study the deconfined phase.   $d_C,d_F$ and $d_S$ are the dimensions of the color, flavor and spin states of $3n+n_p$ quarks, respectively, $m_{FS}$ is the multiplicity of the flavor and spin states, and the summation is taken for all possible states.
Here, we divide it by $n_p$ to normalize the result with respect to the single quark case.  We also divide by $n$ to keep the surrounding baryon at constant density for comparison at the same density.

\subsubsection{Two baryons + one quark}
\label{2b1q}

We now consider a quark around two baryons. The color and flavor $\otimes$ spin states of 7 quarks configuration are as follows.\\
\\
Color : $\begin{tabular}{|c|c|c|}
  \cline{1-3}
  \quad \quad & \quad \quad & \quad \quad  \\
  \cline{1-3}
  \quad \quad & \quad \quad \\
  \cline{1-2}
  \quad \quad & \quad \quad \\
  \cline{1-2}
\end{tabular}$, \quad
Flavor $\otimes$ Spin : $\begin{tabular}{|c|c|c|}
  \cline{1-3}
  \quad \quad & \quad \quad & \quad \quad  \\
  \cline{1-3}
  \quad \quad & \quad \quad & \quad \quad  \\
  \cline{1-3}
  \quad \quad \\
  \cline{1-1}
\end{tabular}.$\\

The flavor $\otimes$ spin coupling state [3,3,1] with SU(6) can be decomposed into the states with the flavor SU(3) and the spin SU(2) as follows.
\begin{widetext}
\begin{align}
[3,3,1]_{FS} &=[6,1]_F \otimes [4,3]_S + [5,2]_F \otimes [5,2]_S + [5,2]_F \otimes [4,3]_S + [5,1,1]_F \otimes [5,2]_S + [5,1,1]_F \otimes [4,3]_S \nonumber\\
 &\quad + [4,3]_F \otimes [6,1]_S + [4,3]_F \otimes [5,2]_S + [4,3]_F \otimes [4,3]_S+ [4,2,1]_F \otimes [6,1]_S + [4,2,1]_F \otimes [5,2]_{S(m=2)} \nonumber\\
  &\quad+ [4,2,1]_F \otimes [4,3]_{S(m=2)} + [3,3,1]_F \otimes [7]_S + [3,3,1]_F \otimes [6,1]_S + [3,3,1]_F \otimes [5,2]_{S(m=2)} \nonumber\\
  &\quad+ [3,3,1]_F \otimes [4,3]_S + [3,2,2]_F \otimes [6,1]_S + [3,2,2]_F \otimes [5,2]_S + [3,2,2]_F \otimes [4,3]_S.
\label{PFS-2b1q}
\end{align}
\end{widetext}

Here we consider the 7 quarks system consisting of two octet baryons and one quark. Then, we can determine the possible flavor states of 7 quarks as follows. \\
Flavor states of 7 quarks :
\begin{align}
\mathbf{8}\times \mathbf{8} \times \mathbf{3} =& \mathbf{3}_{(m=3)}+\overline{\mathbf{6}}_{(m=3)}+\mathbf{15}_{(m=4)}+\mathbf{15'} + \mathbf{24}_{(m=2)} \nonumber\\
&+\mathbf{42}.
\label{PF-2b1q}
\end{align}

Selecting flavor states given in Eq.~\eqref{PF-2b1q} from Eq.~\eqref{PFS-2b1q}, we can represent the possible flavor and spin states as follows.\\

Flavor and spin : $\begin{tabular}{|c|c|c|c}
  \cline{1-3}
  \quad \quad & \quad \quad & \quad \quad \\
  \cline{1-3}
  \quad \quad & \quad \quad \\
  \cline{1-2}
  \quad \quad & \quad \quad \\
  \cline{1-2}
  \multicolumn{4}{c}{$\mathbf{3}(S=\frac{1}{2},\frac{3}{2})$}
\end{tabular}$,
$\begin{tabular}{|c|c|c|c}
  \cline{1-3}
  \quad \quad & \quad \quad & \quad \quad \\
  \cline{1-3}
  \quad \quad & \quad \quad & \quad \quad \\
  \cline{1-3}
  \quad \quad  \\
  \cline{1-1}
  \multicolumn{4}{c}{$\overline{\mathbf{6}}(S=\frac{1}{2},\frac{3}{2})$}
\end{tabular}$,
$\begin{tabular}{|c|c|c|c|c}
  \cline{1-4}
  \quad \quad & \quad \quad & \quad \quad & \quad \quad \\
  \cline{1-4}
  \quad \quad & \quad \quad \\
  \cline{1-2}
  \quad \quad  \\
  \cline{1-1}
  \multicolumn{5}{c}{$\mathbf{15}(S=\frac{1}{2},\frac{3}{2})$}
\end{tabular}$,
$\begin{tabular}{|c|c|c|c|c|c}
  \cline{1-5}
  \quad \quad & \quad \quad & \quad \quad & \quad \quad & \quad \quad \\
  \cline{1-5}
  \quad \quad  \\
  \cline{1-1}
  \quad \quad  \\
  \cline{1-1}
  \multicolumn{6}{c}{$\mathbf{15'}(S=\frac{1}{2},\frac{3}{2})$}
\end{tabular}$\quad,
$\begin{tabular}{|c|c|c|c|c}
  \cline{1-4}
  \quad \quad & \quad \quad & \quad \quad & \quad \quad  \\
  \cline{1-4}
  \quad \quad & \quad \quad & \quad \quad  \\
  \cline{1-3}
  \multicolumn{5}{c}{$\mathbf{24}(S=\frac{1}{2},\frac{3}{2})$}
\end{tabular}$,
$\begin{tabular}{|c|c|c|c|c|c}
  \cline{1-5}
  \quad \quad & \quad \quad & \quad \quad & \quad \quad & \quad \quad  \\
  \cline{1-5}
  \quad \quad & \quad \quad  \\
  \cline{1-2}
  \multicolumn{6}{c}{$\mathbf{42}(S=\frac{1}{2},\frac{3}{2})$}
\end{tabular}$.\\

\subsubsection{Three baryons + one quark}
\label{3b1q}

The color and flavor $\otimes$ spin states of 10 quarks configuration are as follows.\\

Color : $\begin{tabular}{|c|c|c|c|}
  \cline{1-4}
  \quad \quad & \quad \quad & \quad \quad & \quad \quad \\
  \cline{1-4}
  \quad \quad & \quad \quad & \quad \quad \\
  \cline{1-3}
  \quad \quad & \quad \quad & \quad \quad \\
  \cline{1-3}
\end{tabular}$, \quad
Flavor $\otimes$ Spin : $\begin{tabular}{|c|c|c|}
  \cline{1-3}
  \quad \quad & \quad \quad & \quad \quad  \\
  \cline{1-3}
  \quad \quad & \quad \quad & \quad \quad  \\
  \cline{1-3}
  \quad \quad & \quad \quad & \quad \quad  \\
  \cline{1-3}
  \quad \quad \\
  \cline{1-1}
\end{tabular}.$\\
The flavor $\otimes$ spin coupling state [3,3,3,1] with SU(6) can be decomposed into the states with the flavor SU(3) and the spin SU(2) as follows.
\begin{widetext}
\begin{align}
[3,3,3,1]_{FS} &=[6,4]_F \otimes [6,4]_S + [6,3,1]_F \otimes [7,3]_S + [6,3,1]_F \otimes [6,4]_S + [6,3,1]_F \otimes [5,5]_S + [6,2,2]_F \otimes [6,4]_S \nonumber\\
  &\quad + [5,5]_F \otimes [5,5]_S + [5,4,1]_F \otimes [7,3]_S + [5,4,1]_F \otimes [6,4]_{S(m=2)} + [5,4,1]_F \otimes [5,5]_S  \nonumber\\
  &\quad + [5,3,2]_F \otimes [8,2]_S + [5,3,2]_F \otimes [7,3]_{S(m=2)} + [5,3,2]_F \otimes [6,4]_{S(m=2)} + [5,3,2]_F \otimes [5,5]_S \nonumber\\
   &\quad + [4,4,2]_F \otimes [8,2]_S + [4,4,2]_F \otimes [7,3]_S + [4,4,2]_F \otimes [6,4]_{S(m=2)} + [4,3,3]_F \otimes [9,1]_S \nonumber\\
    &\quad + [4,3,3]_F \otimes [8,2]_S + [4,3,3]_F \otimes [7,3]_{S(m=2)} + [4,3,3]_F \otimes [6,4]_S + [4,3,3]_F \otimes [5,5]_S.
\end{align}
\end{widetext}
Similar to the two baryons and one quark case, we can determine the possible flavor states of 10 quarks as follows.\\

Flavor states of 10 quarks :
\begin{align}
  \mathbf{8}\times &\mathbf{8} \times \mathbf{8} \times
  \mathbf{3} \nonumber\\ 
  = & \mathbf{3}_{(m=10)}+\overline{\mathbf{6}}_{(m=12)}+\mathbf{15}_{(m=18)} +\mathbf{15'}_{(m=6)} +\overline{\mathbf{21}}_{(m=2)} \nonumber\\
  &+\overline{\mathbf{24}}_{(m=12)}+\mathbf{42}_{(m=9)}+\mathbf{48}_{(m=2)}+\overline{\mathbf{60}}_{(m=3)} +\mathbf{90}.
\end{align}

Combining the two equation above, we find the following possible states.

Flavor and spin : $\begin{tabular}{|c|c|c|c|c}
  \cline{1-4}
  \quad \quad & \quad \quad & \quad \quad & \quad \quad \\
  \cline{1-4}
  \quad \quad & \quad \quad & \quad \quad \\
  \cline{1-3}
  \quad \quad & \quad \quad & \quad \quad \\
  \cline{1-3}
  \multicolumn{5}{c}{$\mathbf{3}(S=0,1,2)$}
\end{tabular}$,
$\begin{tabular}{|c|c|c|c|c}
  \cline{1-4}
  \quad \quad & \quad \quad & \quad \quad & \quad \quad \\
  \cline{1-4}
  \quad \quad & \quad \quad & \quad \quad & \quad \quad \\
  \cline{1-4}
  \quad \quad & \quad \quad  \\
  \cline{1-2}
  \multicolumn{5}{c}{$\mathbf{\overline{6}}(S=0,1,2)$}
\end{tabular}$,
$\begin{tabular}{|c|c|c|c|c|c}
  \cline{1-5}
  \quad \quad & \quad \quad & \quad \quad & \quad \quad & \quad \quad \\
  \cline{1-5}
  \quad \quad & \quad \quad & \quad \quad \\
  \cline{1-3}
  \quad \quad & \quad \quad  \\
  \cline{1-2}
  \multicolumn{6}{c}{$\mathbf{15}(S=0,1,2)$}
\end{tabular}$,
$\begin{tabular}{|c|c|c|c|c|c|c}
  \cline{1-6}
  \quad \quad & \quad \quad & \quad \quad & \quad \quad & \quad \quad & \quad \quad \\
  \cline{1-6}
  \quad \quad & \quad \quad  \\
  \cline{1-2}
  \quad \quad & \quad \quad  \\
  \cline{1-2}
  \multicolumn{7}{c}{$\mathbf{15'}(S=1)$}
\end{tabular}$ \qquad \quad ,
$\begin{tabular}{|c|c|c|c|c|c}
  \cline{1-5}
  \quad \quad & \quad \quad & \quad \quad & \quad \quad & \quad \quad  \\
  \cline{1-5}
  \quad \quad & \quad \quad & \quad \quad & \quad \quad & \quad \quad  \\
  \cline{1-5}
  \multicolumn{6}{c}{$\mathbf{\overline{21}}(S=0)$}
\end{tabular}$ \quad \quad ,
$\begin{tabular}{|c|c|c|c|c|c}
  \cline{1-5}
  \quad \quad & \quad \quad & \quad \quad & \quad \quad & \quad \quad  \\
  \cline{1-5}
  \quad \quad & \quad \quad & \quad \quad & \quad \quad  \\
  \cline{1-4}
  \quad \quad \\
  \cline{1-1}
  \multicolumn{6}{c}{$\mathbf{\overline{24}}(S=0,1,2)$}
\end{tabular}$,
$\begin{tabular}{|c|c|c|c|c|c|c}
  \cline{1-6}
  \quad \quad & \quad \quad & \quad \quad & \quad \quad & \quad \quad & \quad \quad  \\
  \cline{1-6}
  \quad \quad & \quad \quad & \quad \quad  \\
  \cline{1-3}
  \quad \quad \\
  \cline{1-1}
  \multicolumn{7}{c}{$\mathbf{42}(S=0,1,2)$}
\end{tabular}$ \quad \quad,
$\begin{tabular}{|c|c|c|c|c|c|c}
  \cline{1-6}
  \quad \quad & \quad \quad & \quad \quad & \quad \quad & \quad \quad & \quad \quad  \\
  \cline{1-6}
  \quad \quad & \quad \quad & \quad \quad & \quad \quad \\
  \cline{1-4}
  \multicolumn{7}{c}{$\mathbf{\overline{60}}(S=1)$}
\end{tabular}\qquad .$\\

\subsubsection{Four baryons + one quark}
\label{4b1q}

The color and flavor $\otimes$ spin states of 13 quarks configuration are as follows.\\

Color : $\begin{tabular}{|c|c|c|c|c|}
  \hline
  \quad \quad & \quad \quad & \quad \quad & \quad \quad & \quad \quad \\
  \hline
  \quad \quad & \quad \quad & \quad \quad & \quad \quad \\
  \cline{1-4}
  \quad \quad & \quad \quad & \quad \quad & \quad \quad \\
  \cline{1-4}
\end{tabular}$, \quad
Flavor $\otimes$ Spin : $\begin{tabular}{|c|c|c|}
  \hline
  \quad \quad & \quad \quad & \quad \quad  \\
  \hline
  \quad \quad & \quad \quad & \quad \quad  \\
  \hline
  \quad \quad & \quad \quad & \quad \quad  \\
  \hline
  \quad \quad & \quad \quad & \quad \quad  \\
  \hline
  \quad \quad \\
  \cline{1-1}
\end{tabular}.$\\
The flavor $\otimes$ spin coupling state [3,3,3,3,1] with SU(6) can be decomposed into the states with the flavor SU(3) and the spin SU(2) as follows.
\begin{widetext}
\begin{align}
[3,3,3,3,1]_{FS} &=[6,6,1]_F \otimes [7,6]_S + [6,5,2]_F \otimes [8,5]_S + [6,5,2]_F \otimes [7,6]_S + [6,4,3]_F \otimes [9,4]_S + [6,4,3]_F \otimes [8,5]_S \nonumber\\
  &\quad + [6,4,3]_F \otimes [7,6]_S + [5,5,3]_F \otimes [8,5]_S + [5,5,3]_F \otimes [7,6]_S + [5,4,4]_F \otimes [8,5]_S + [5,4,4]_F \otimes [7,6]_S
\end{align}
Flavor states of 13 quarks :
\begin{align}
  \mathbf{8}\times \mathbf{8} \times \mathbf{8} \times \mathbf{8} \times \mathbf{3}
  =&\mathbf{3}_{(m=40)}+\overline{\mathbf{6}}_{(m=52)}+\mathbf{15}_{(m=85)}+\mathbf{15'}_{(m=35)}+\mathbf{21}_{(m=17)}+\mathbf{24}_{(m=68)}
+\mathbf{36}_{(m=2)}+\mathbf{42}_{(m=60)} \nonumber\\
&+\mathbf{48}_{(m=20)}+\overline{\mathbf{60}}_{(m=30)}+\mathbf{63}_{(m=5)}+\mathbf{90}_{(m=16)} + +\mathbf{105}_{(m=3)}+\mathbf{120}_{(m=4)}+\mathbf{165}.
\end{align}
Flavor and spin : $\begin{tabular}{|c|c|c|c|c|c}
  \cline{1-5}
  \quad \quad & \quad \quad & \quad \quad & \quad \quad & \quad \quad \\
  \cline{1-5}
  \quad \quad & \quad \quad & \quad \quad & \quad \quad \\
  \cline{1-4}
  \quad \quad & \quad \quad & \quad \quad & \quad \quad\\
  \cline{1-4}
  \multicolumn{5}{c}{$\mathbf{3}(S=\frac{1}{2},\frac{3}{2})$}
\end{tabular}$,
$\begin{tabular}{|c|c|c|c|c|c}
  \cline{1-5}
  \quad \quad & \quad \quad & \quad \quad & \quad \quad & \quad \quad \\
  \cline{1-5}
  \quad \quad & \quad \quad & \quad \quad & \quad \quad & \quad \quad \\
  \cline{1-5}
  \quad \quad & \quad \quad & \quad \quad \\
  \cline{1-3}
  \multicolumn{5}{c}{$\mathbf{\overline{6}}(S=\frac{1}{2},\frac{3}{2})$}
\end{tabular}$,
$\begin{tabular}{|c|c|c|c|c|c|c}
  \cline{1-6}
  \quad \quad & \quad \quad & \quad \quad & \quad \quad & \quad \quad & \quad \quad\\
  \cline{1-6}
  \quad \quad & \quad \quad & \quad \quad & \quad \quad \\
  \cline{1-4}
  \quad \quad & \quad \quad & \quad \quad \\
  \cline{1-3}
  \multicolumn{6}{c}{$\mathbf{15}(S=\frac{1}{2},\frac{3}{2},\frac{5}{2})$}
\end{tabular}$,
$\begin{tabular}{|c|c|c|c|c|c|c}
  \cline{1-6}
  \quad \quad & \quad \quad & \quad \quad & \quad \quad & \quad \quad & \quad \quad \\
  \cline{1-6}
  \quad \quad & \quad \quad & \quad \quad & \quad \quad & \quad \quad & \quad \quad \\
  \cline{1-6}
  \quad \quad \\
  \cline{1-1}
  \multicolumn{6}{c}{$\mathbf{21}(S=\frac{1}{2})$}
\end{tabular}$,
$\begin{tabular}{|c|c|c|c|c|c|c}
  \cline{1-6}
  \quad \quad & \quad \quad & \quad \quad & \quad \quad & \quad \quad & \quad \quad \\
  \cline{1-6}
  \quad \quad & \quad \quad & \quad \quad & \quad \quad & \quad \quad \\
  \cline{1-5}
  \quad \quad & \quad \quad \\
  \cline{1-2}
  \multicolumn{6}{c}{$\mathbf{24}(S=\frac{1}{2},\frac{3}{2})$}
\end{tabular}.$\\
\end{widetext}

\subsubsection{Five baryons + one quark}
\label{5b1q}

The color and flavor $\otimes$ spin states of 16 quarks configuration are as follows.\\
\begin{small}
Color : $\begin{tabular}{|c|c|c|c|c|c|}
  \hline
  \quad \quad & \quad \quad & \quad \quad & \quad \quad & \quad \quad & \quad \quad \\
  \hline
  \quad \quad & \quad \quad & \quad \quad & \quad \quad & \quad \quad \\
  \cline{1-5}
  \quad \quad & \quad \quad & \quad \quad & \quad \quad & \quad \quad \\
  \cline{1-5}
\end{tabular}$, \
Flavor $\otimes$ Spin : $\begin{tabular}{|c|c|c|}
  \hline
  \quad \quad & \quad \quad & \quad \quad  \\
  \hline
  \quad \quad & \quad \quad & \quad \quad  \\
  \hline
  \quad \quad & \quad \quad & \quad \quad  \\
  \hline
  \quad \quad & \quad \quad & \quad \quad  \\
  \hline
  \quad \quad & \quad \quad & \quad \quad  \\
  \hline
  \quad \quad \\
  \cline{1-1}
\end{tabular}.$
\end{small}\\

The flavor $\otimes$ spin coupling state [3,3,3,3,3,1] with SU(6) can be decomposed into the states with the flavor SU(3) and the spin SU(2) as follows.
\begin{align}
[3,3,3,3,3,1]_{FS} &=[6,6,4]_F \otimes [9,7]_S + [6,5,5]_F \otimes [8,8]_S.
\end{align}

\begin{widetext}
Flavor states of 16 quarks :
\begin{align}
  \mathbf{8}\times \mathbf{8}\times \mathbf{8} \times \mathbf{8} \times \mathbf{8} \times \mathbf{3}
  =&\mathbf{3}_{(m=177)}+\overline{\mathbf{6}}_{(m=245)}+\mathbf{15}_{(m=425)}+\mathbf{15'}_{(m=200)}+\mathbf{21}_{(m=120)}+\mathbf{24}_{(m=380)}
+\mathbf{36}_{(m=25)} \nonumber\\
& +\mathbf{42}_{(m=374)} +\mathbf{45}_{(m=5)}+\mathbf{48}_{(m=156)}+\mathbf{60}_{(m=230)}+\mathbf{63}_{(m=61)}+\mathbf{90}_{(m=150)}+\mathbf{99}_{(m=5)} \nonumber\\
& +\mathbf{105}_{(m=45)}+\mathbf{120}_{(m=60)} +\mathbf{132}_{(m=9)}+\mathbf{165}_{(m=25)}+\mathbf{192}_{(m=4)}+\mathbf{210}_{(m=5)}+\mathbf{273}.
\end{align}

Flavor and spin : $\begin{tabular}{|c|c|c|c|c|c|}
  \hline
  \quad \quad & \quad \quad & \quad \quad & \quad \quad & \quad \quad & \quad \quad \\
  \hline
  \quad \quad & \quad \quad & \quad \quad & \quad \quad & \quad \quad \\
  \cline{1-5}
  \quad \quad & \quad \quad & \quad \quad & \quad \quad & \quad \quad \\
  \cline{1-5}
  \multicolumn{5}{c}{$\mathbf{3}(S=0)$}
\end{tabular}$,
$\begin{tabular}{|c|c|c|c|c|c|}
  \hline
  \quad \quad & \quad \quad & \quad \quad & \quad \quad & \quad \quad & \quad \quad \\
  \hline
  \quad \quad & \quad \quad & \quad \quad & \quad \quad & \quad \quad & \quad \quad \\
  \hline
  \quad \quad & \quad \quad & \quad \quad & \quad \quad \\
  \cline{1-4}
  \multicolumn{6}{c}{$\mathbf{\overline{6}}(S=1)$}
\end{tabular}$.\\
\end{widetext}

\subsection{Diquark case($C=\overline{\mathbf{3}},F=\overline{\mathbf{3}},S=0$)}
\label{diquark_AAA}

From Sec.\ref{diquark_AAA} to Sec.\ref{diquark_SSA}, we construct the multiquark state containing all possible diquarks as a probe. There are four possible color-flavor-spin states of diquark satisfying the Pauli principle. Among them, in this section, we construct the multiquark state containing the most stable diquark.

\subsubsection{one baryon + one diquark}

\label{1b1d}
Let us consider the multiquark system consisting of one baryon and one diquark.  The color state of five quarks should be a antitriplet. Then, the flavor $\otimes$ spin coupling state of five quarks can be determined as follows. \\

Color : $\begin{tabular}{|c|c|}
  \cline{1-2}
  \quad \quad & \quad \quad   \\
  \cline{1-2}
  \quad \quad & \quad \quad \\
  \cline{1-2}
  \quad \quad \\
  \cline{1-1}
\end{tabular}$, \quad
Flavor $\otimes$ Spin : $\begin{tabular}{|c|c|c|}
  \cline{1-3}
  \quad \quad & \quad \quad & \quad \quad  \\
  \cline{1-3}
  \quad \quad & \quad \quad \\
  \cline{1-2}
\end{tabular}.$\\

We can decompose SU(6) flavor $\otimes$ spin coupling state into SU(3) flavor and SU(2) spin states as follows.
\begin{widetext}
\begin{align}
  [3,2]_{FS} &= [5]_F \otimes [3,2]_S + [4,1]_F \otimes [4,1]_S + [4,1]_F \otimes [3,2]_S + [3,2]_F \otimes [5]_S + [3,2]_F \otimes [4,1]_S + [3,2]_F \otimes [3,2]_S  \nonumber\\
   & \quad + [3,1,1]_F \otimes [4,1]_S + [3,1,1]_F \otimes [3,2]_S + [2,2,1]_F \otimes [4,1]_S + [2,2,1]_F \otimes [3,2]_S.
\end{align}
\end{widetext}

Now, we can classify all possible flavor and spin states of a five quarks configuration as follows.\\

Flavor states of 5 quarks :
\begin{align}
  \mathbf{8}\times \overline{\mathbf{3}}=\overline{\mathbf{3}}+\mathbf{6}+\overline{\mathbf{15}}.
\end{align}

Flavor and spin : $\begin{tabular}{|c|c|c}
  \cline{1-2}
  \quad \quad & \quad \quad  \\
  \cline{1-2}
  \quad \quad & \quad \quad \\
  \cline{1-2}
  \quad \quad \\
  \cline{1-1}
  \multicolumn{3}{c}{$\overline{\mathbf{3}}(S=\frac{1}{2})$}
\end{tabular}$,
$\begin{tabular}{|c|c|c|c}
  \cline{1-3}
  \quad \quad & \quad \quad & \quad \quad  \\
  \cline{1-3}
  \quad \quad \\
  \cline{1-1}
  \quad \quad \\
  \cline{1-1}
  \multicolumn{4}{c}{$\mathbf{6}(S=\frac{1}{2})$}
\end{tabular}$,
$\begin{tabular}{|c|c|c|c}
  \cline{1-3}
  \quad \quad & \quad \quad & \quad \quad  \\
  \cline{1-3}
  \quad \quad & \quad \quad\\
  \cline{1-2}
  \multicolumn{4}{c}{$\overline{\mathbf{15}}(S=\frac{1}{2})$}
\end{tabular}$.\\

\subsubsection{two baryons + one diquark}
\label{2b1d}

The color state of 8 quarks should be a antitriplet. Then, the flavor $\otimes$ spin coupling state of eight quarks should be its conjugate as follows. \\

Color : $\begin{tabular}{|c|c|c|}
  \cline{1-3}
  \quad \quad & \quad \quad & \quad \quad  \\
  \cline{1-3}
  \quad \quad & \quad \quad & \quad \quad \\
  \cline{1-3}
  \quad \quad & \quad \quad \\
  \cline{1-2}
\end{tabular}$, \quad
Flavor $\otimes$ Spin : $\begin{tabular}{|c|c|c|}
  \cline{1-3}
  \quad \quad & \quad \quad & \quad \quad  \\
  \cline{1-3}
  \quad \quad & \quad \quad & \quad \quad \\
  \cline{1-3}
  \quad \quad & \quad \quad \\
  \cline{1-2}
\end{tabular}.$\\

We can decompose SU(6) flavor $\otimes$ spin coupling state into flavor SU(3) and spin SU(2) states as follows.
\begin{widetext}
\begin{align}
  [3,3,2]_{FS} &= [6,2]_F \otimes [5,3]_S + [6,1,1]_F \otimes [4,4]_S + [5,3]_F \otimes [6,2]_S + [5,3]_F \otimes [5,3]_S + [5,3]_F \otimes [4,4]_S + [5,2,1]_F \otimes [6,2]_S \nonumber\\
   & \quad  + [5,2,1]_F \otimes [5,3]_{S(m=2)} + [5,2,1]_F \otimes [4,4]_S + [4,4]_F \otimes [5,3]_S + [4,3,1]_F \otimes [7,1]_S + [4,3,1]_F \otimes [6,2]_{S(m=2)} \nonumber\\
    &\quad + [4,3,1]_F \otimes [5,3]_{S(m=2)} + [4,3,1]_F \otimes [4,4]_S + [4,2,2]_F \otimes [7,1]_S + [4,2,2]_F \otimes [6,2]_S + [4,2,2]_F \otimes [5,3]_{S(m=2)} \nonumber\\
    &\quad + [3,3,2]_F \otimes [8]_S + [3,3,2]_F \otimes [7,1]_S + [3,3,2]_F \otimes [6,2]_S + [3,3,2]_F \otimes [5,3]_S + [3,3,2]_F \otimes [4,4]_S.
\end{align}

Now, we can classify all possible flavor and spin states of 8 quarks configuration as follows.\\

Flavor states of 8 quarks :
\begin{align}
  \mathbf{8}\times \mathbf{8}\times \overline{\mathbf{3}} &=\overline{\mathbf{3}}_{(m=3)} + \mathbf{6}_{(m=3)} + \overline{\mathbf{15}}_{(m=4)} + \overline{\mathbf{15'}}_{(m=2)} + \overline{\mathbf{24}}_{(m=2)} + \overline{\mathbf{42}}.
\end{align}

Flavor and spin : $\begin{tabular}{|c|c|c|c}
  \cline{1-3}
  \quad \quad & \quad \quad & \quad \quad \\
  \cline{1-3}
  \quad \quad & \quad \quad & \quad \quad \\
  \cline{1-3}
  \quad \quad & \quad \quad \\
  \cline{1-2}
  \multicolumn{4}{c}{$\overline{\mathbf{3}}(S=0,1)$}
\end{tabular}$,
$\begin{tabular}{|c|c|c|c|c}
  \cline{1-4}
  \quad \quad & \quad \quad & \quad \quad & \quad \quad  \\
  \cline{1-4}
  \quad \quad & \quad \quad \\
  \cline{1-2}
  \quad \quad & \quad \quad \\
  \cline{1-2}
  \multicolumn{5}{c}{$\mathbf{6}(S=1)$}\quad \
\end{tabular}$,
$\begin{tabular}{|c|c|c|c|c}
  \cline{1-4}
  \quad \quad & \quad \quad & \quad \quad & \quad \quad  \\
  \cline{1-4}
  \quad \quad & \quad \quad & \quad \quad \\
  \cline{1-3}
  \quad \quad \\
  \cline{1-1}
  \multicolumn{5}{c}{$\overline{\mathbf{15}}(S=0,1)$}
\end{tabular}$,
$\begin{tabular}{|c|c|c|c|}
  \cline{1-4}
  \quad \quad & \quad \quad & \quad \quad & \quad \quad  \\
  \cline{1-4}
  \quad \quad & \quad \quad & \quad \quad & \quad \quad \\
  \cline{1-4}
  \multicolumn{4}{c}{$\overline{\mathbf{15'}}(S=1)$}
\end{tabular}$\ ,
$\begin{tabular}{|c|c|c|c|c|c}
  \cline{1-5}
  \quad \quad & \quad \quad & \quad \quad & \quad \quad & \quad \quad \\
  \cline{1-5}
  \quad \quad & \quad \quad \\
  \cline{1-2}
  \quad \quad \\
  \cline{1-1}
  \multicolumn{6}{c}{$\overline{\mathbf{24}}(S=0,1)$}\quad \
\end{tabular}$,
$\begin{tabular}{|c|c|c|c|c|c}
  \cline{1-5}
  \quad \quad & \quad \quad & \quad \quad & \quad \quad & \quad \quad \\
  \cline{1-5}
  \quad \quad & \quad \quad & \quad \quad \\
  \cline{1-3}
  \multicolumn{6}{c}{$\overline{\mathbf{42}}(S=0,1)$}\quad \
\end{tabular}$.\\
\end{widetext}

\subsubsection{three baryons + one diquark}
\label{3b1d}

The color and flavor $\otimes$ spin states of 11 quarks configuration are as follows.\\

Color : $\begin{tabular}{|c|c|c|c|}
  \cline{1-4}
  \quad \quad & \quad \quad & \quad \quad & \quad \quad \\
  \cline{1-4}
  \quad \quad & \quad \quad & \quad \quad & \quad \quad \\
  \cline{1-4}
  \quad \quad & \quad \quad & \quad \quad \\
  \cline{1-3}
\end{tabular}$, \quad
Flavor $\otimes$ Spin : $\begin{tabular}{|c|c|c|}
  \cline{1-3}
  \quad \quad & \quad \quad & \quad \quad  \\
  \cline{1-3}
  \quad \quad & \quad \quad & \quad \quad \\
  \cline{1-3}
  \quad \quad & \quad \quad & \quad \quad \\
  \cline{1-3}
  \quad \quad & \quad \quad \\
  \cline{1-2}
\end{tabular}.$\\

We can decompose flavor $\otimes$ spin coupling SU(6)  state into flavor SU(3) and spin SU(2) states as follows.
\begin{widetext}
\begin{align}
  [3,3,3,2]_{FS} &= [6,5]_F \otimes [6,5]_S + [6,4,1]_F \otimes [7,4]_S + [6,4,1]_F \otimes [6,5]_S + [6,3,2]_F \otimes [8,3]_S + [6,3,2]_F \otimes [7,4]_S  \nonumber\\
   & \quad + [6,3,2]_F \otimes [6,5]_S  + [5,5,1]_F \otimes [7,4]_S + [5,5,1]_F \otimes [6,5]_S + [5,4,2]_F \otimes [8,3]_S  \nonumber\\
    &\quad + [5,4,2]_F \otimes [7,4]_{S(m=2)} + [5,4,2]_F \otimes [6,5]_{S(m=2)} + [5,3,3]_F \otimes [9,2]_S + [5,3,3]_F \otimes [8,3]_S  \nonumber\\
    &\quad + [5,3,3]_F \otimes [7,4]_{S(m=2)} + [5,3,3]_F \otimes [6,5]_S + [4,4,3]_F \otimes [8,3]_S + [4,4,3]_F \otimes [7,4]_S \nonumber\\
    &\quad + [4,4,3]_F \otimes [6,5]_S.
\end{align}

Now, we can classify all possible flavor and spin states of a 11 quarks configuration as follows.\\
\\
Flavor states of 11 quarks :
\begin{align}
\mathbf{8}\times \mathbf{8}\times \mathbf{8}\times \overline{\mathbf{3}} =& \overline{\mathbf{3}}_{(m=10)} + \mathbf{6}_{(m=12)} + \overline{\mathbf{15}}_{(m=18)} + \overline{\mathbf{15'}}_{(m=6)} + \overline{\mathbf{24}}_{(m=12)} + \overline{\mathbf{42}}_{(m=9)} + \overline{\mathbf{21}}_{(m=2)} + \overline{\mathbf{60}}_{(m=3)} \nonumber \\
&  + \overline{\mathbf{48}}_{(m=2)} + \overline{\mathbf{90}}.
\end{align}

Flavor and spin : $\begin{tabular}{|c|c|c|c|c}
  \cline{1-4}
  \quad \quad & \quad \quad & \quad \quad & \quad \quad \\
  \cline{1-4}
  \quad \quad & \quad \quad & \quad \quad & \quad \quad \\
  \cline{1-4}
  \quad \quad & \quad \quad & \quad \quad \\
  \cline{1-3}
  \multicolumn{5}{c}{$\overline{\mathbf{3}}(S=\frac{1}{2},\frac{3}{2})$}
\end{tabular}$,
$\begin{tabular}{|c|c|c|c|c|c}
  \cline{1-5}
  \quad \quad & \quad \quad & \quad \quad & \quad \quad & \quad \quad \\
  \cline{1-5}
  \quad \quad & \quad \quad & \quad \quad \\
  \cline{1-3}
  \quad \quad & \quad \quad & \quad \quad \\
  \cline{1-3}
  \multicolumn{6}{c}{$\mathbf{6}(S=\frac{1}{2},\frac{3}{2})$}
\end{tabular}$\qquad ,
$\begin{tabular}{|c|c|c|c|c|c}
  \cline{1-5}
  \quad \quad & \quad \quad & \quad \quad & \quad \quad & \quad \quad  \\
  \cline{1-5}
  \quad \quad & \quad \quad & \quad \quad & \quad \quad \\
  \cline{1-4}
  \quad \quad & \quad \quad \\
  \cline{1-2}
  \multicolumn{6}{c}{$\overline{\mathbf{15}}(S=\frac{1}{2},\frac{3}{2})$}
\end{tabular}$\quad ,
$\begin{tabular}{|c|c|c|c|c|c}
  \cline{1-5}
  \quad \quad & \quad \quad & \quad \quad & \quad \quad & \quad \quad  \\
  \cline{1-5}
  \quad \quad & \quad \quad & \quad \quad & \quad \quad & \quad \quad \\
  \cline{1-5}
  \quad \quad \\
  \cline{1-1}
  \multicolumn{6}{c}{$\overline{\mathbf{15'}}(S=\frac{1}{2},\frac{3}{2})$}
\end{tabular}$\ , \
$\begin{tabular}{|c|c|c|c|c|c|c}
  \cline{1-6}
  \quad \quad & \quad \quad & \quad \quad & \quad \quad & \quad \quad & \quad \quad \\
  \cline{1-6}
  \quad \quad & \quad \quad & \quad \quad \\
  \cline{1-3}
  \quad \quad & \quad \quad \\
  \cline{1-2}
  \multicolumn{7}{c}{$\overline{\mathbf{24}}(S=\frac{1}{2},\frac{3}{2})$}
\end{tabular}$\quad \quad,\\
$\begin{tabular}{|c|c|c|c|c|c|c}
  \cline{1-6}
  \quad \quad & \quad \quad & \quad \quad & \quad \quad & \quad \quad & \quad \quad \\
  \cline{1-6}
  \quad \quad & \quad \quad & \quad \quad & \quad \quad \\
  \cline{1-4}
  \quad \quad \\
  \cline{1-1}
  \multicolumn{7}{c}{$\overline{\mathbf{42}}(S=\frac{1}{2},\frac{3}{2})$}
\end{tabular}$ \qquad \quad ,
$\begin{tabular}{|c|c|c|c|c|c|c}
  \cline{1-6}
  \quad \quad & \quad \quad & \quad \quad & \quad \quad & \quad \quad & \quad \quad \\
  \cline{1-6}
  \quad \quad & \quad \quad & \quad \quad & \quad \quad & \quad \quad \\
  \cline{1-5}
  \multicolumn{6}{c}{$\overline{\mathbf{48}}(S=\frac{1}{2})$}
\end{tabular}$.\\
\end{widetext}

\subsubsection{Four baryons + one diquark}
\label{4b1d}

The color and flavor $\otimes$ spin states of 14 quarks configuration are as follows.\\

\begin{small}
Color : $\begin{tabular}{|c|c|c|c|c|}
  \hline
  \quad \quad & \quad \quad & \quad \quad & \quad \quad & \quad \quad \\
  \hline
  \quad \quad & \quad \quad & \quad \quad & \quad \quad & \quad \quad \\
  \cline{1-5}
  \quad \quad & \quad \quad & \quad \quad & \quad \quad \\
  \cline{1-4}
\end{tabular}$, \quad
Flavor $\otimes$ Spin : $\begin{tabular}{|c|c|c|}
  \hline
  \quad \quad & \quad \quad & \quad \quad  \\
  \hline
  \quad \quad & \quad \quad & \quad \quad  \\
  \hline
  \quad \quad & \quad \quad & \quad \quad  \\
  \hline
  \quad \quad & \quad \quad & \quad \quad  \\
  \hline
  \quad \quad & \quad \quad \\
  \cline{1-2}
\end{tabular}.$
\end{small}\\
The flavor $\otimes$ spin coupling state [3,3,3,3,2] with SU(6) can be decomposed into the states with the flavor SU(3) and the spin SU(2) as follows.

\begin{widetext}
\begin{align}
[3,3,3,3,2]_{FS} &=[6,6,2]_F \otimes [8,6]_S + [6,5,3]_F \otimes [9,5]_S + [6,5,3]_F \otimes [8,6]_S + [6,5,3]_F \otimes [7,7]_S + [6,4,4]_F \otimes [8,6]_S \nonumber\\
  &\quad + [5,5,4]_F \otimes [8,6]_S + [5,5,4]_F \otimes [7,7]_S.
\end{align}

Flavor states of 14 quarks :
\begin{align}
  \mathbf{8}\times \mathbf{8} \times \mathbf{8} \times \mathbf{8} \times \overline{\mathbf{3}}
  =&\overline{\mathbf{3}}_{(m=40)}+\mathbf{6}_{(m=52)}+\overline{\mathbf{15}}_{(m=85)}+\overline{\mathbf{15'}}_{(m=35)}+\overline{\mathbf{21}}_{(m=17)}+\overline{\mathbf{24}}_{(m=68)}
+\overline{\mathbf{36}}_{(m=2)}+\overline{\mathbf{42}}_{(m=60)} \nonumber\\
&+\overline{\mathbf{48}}_{(m=20)}+\overline{\mathbf{60}}_{(m=30)}+\overline{\mathbf{63}}_{(m=5)}+\overline{\mathbf{90}}_{(m=16)} + +\overline{\mathbf{105}}_{(m=3)}+\overline{\mathbf{120}}_{(m=4)}+\overline{\mathbf{165}}.
\end{align}

Flavor and spin : $\begin{tabular}{|c|c|c|c|c|c}
  \hline
  \quad \quad & \quad \quad & \quad \quad & \quad \quad & \quad \quad \\
  \hline
  \quad \quad & \quad \quad & \quad \quad & \quad \quad & \quad \quad \\
  \hline
  \quad \quad & \quad \quad & \quad \quad & \quad \quad\\
  \cline{1-4}
  \multicolumn{5}{c}{$\overline{\mathbf{3}}(S=0,1)$}
\end{tabular}$,
$\begin{tabular}{|c|c|c|c|c|c|}
  \hline
  \quad \quad & \quad \quad & \quad \quad & \quad \quad & \quad \quad & \quad \quad \\
  \hline
  \quad \quad & \quad \quad & \quad \quad & \quad \quad \\
  \cline{1-4}
  \quad \quad & \quad \quad & \quad \quad & \quad \quad \\
  \cline{1-4}
  \multicolumn{5}{c}{$\mathbf{6}(S=1)$}
\end{tabular}$,
$\begin{tabular}{|c|c|c|c|c|c|}
  \hline
  \quad \quad & \quad \quad & \quad \quad & \quad \quad & \quad \quad & \quad \quad\\
  \hline
  \quad \quad & \quad \quad & \quad \quad & \quad \quad & \quad \quad \\
  \cline{1-5}
  \quad \quad & \quad \quad & \quad \quad \\
  \cline{1-3}
  \multicolumn{6}{c}{$\overline{\mathbf{15}}(S=0,1,2)$}
\end{tabular}$,
$\begin{tabular}{|c|c|c|c|c|c|}
  \hline
  \quad \quad & \quad \quad & \quad \quad & \quad \quad & \quad \quad & \quad \quad \\
  \hline
  \quad \quad & \quad \quad & \quad \quad & \quad \quad & \quad \quad & \quad \quad \\
  \hline
  \quad \quad & \quad \quad \\
  \cline{1-2}
  \multicolumn{6}{c}{$\overline{\mathbf{15'}}(S=1)$}
\end{tabular}$.
\end{widetext}

\subsubsection{Five baryons + one diquark}
\label{5b1d}

The color and flavor $\otimes$ spin states of 17 quarks configuration are as follows.\\

\begin{small}
Color : $\begin{tabular}{|c|c|c|c|c|c|}
  \hline
  \quad \quad & \quad \quad & \quad \quad & \quad \quad & \quad \quad & \quad \quad \\
  \hline
  \quad \quad & \quad \quad & \quad \quad & \quad \quad & \quad \quad & \quad \quad \\
  \hline
  \quad \quad & \quad \quad & \quad \quad & \quad \quad & \quad \quad \\
  \cline{1-5}
\end{tabular}$, 
Flavor $\otimes$ Spin : $\begin{tabular}{|c|c|c|}
  \hline
  \quad \quad & \quad \quad & \quad \quad  \\
  \hline
  \quad \quad & \quad \quad & \quad \quad  \\
  \hline
  \quad \quad & \quad \quad & \quad \quad  \\
  \hline
  \quad \quad & \quad \quad & \quad \quad  \\
  \hline
  \quad \quad & \quad \quad & \quad \quad  \\
  \hline
  \quad \quad & \quad \quad \\
  \cline{1-2}
\end{tabular}.$
\end{small}\\
The flavor $\otimes$ spin coupling state [3,3,3,3,3,2] with SU(6) can be decomposed into the states with the flavor SU(3) and the spin SU(2) as follows.
\begin{align}
[3,3,3,3,3,2]_{FS} &=[6,6,5]_F \otimes [9,8]_S.
\end{align}

\begin{widetext}
Flavor states of 17 quarks :
\begin{align}
  \mathbf{8}\times \mathbf{8}\times \mathbf{8} \times \mathbf{8} \times \mathbf{8} \times \overline{\mathbf{3}}
  =& \overline{\mathbf{3}}_{(m=177)}+\mathbf{6}_{(m=245)}+\overline{\mathbf{15}}_{(m=425)}+\overline{\mathbf{15'}}_{(m=200)}+ \overline{\mathbf{21}}_{(m=120)}+\overline{\mathbf{24}}_{(m=380)}
+ \overline{\mathbf{36}}_{(m=25)} \nonumber\\
& +\overline{\mathbf{42}}_{(m=374)} + \overline{\mathbf{45}}_{(m=5)}+ \overline{\mathbf{48}}_{(m=156)} +\overline{\mathbf{60}}_{(m=230)} +\overline{\mathbf{63}}_{(m=61)} +\overline{\mathbf{90}}_{(m=150)} +\overline{\mathbf{99}}_{(m=5)}  \nonumber\\
& +\overline{\mathbf{105}}_{(m=45)} +\overline{\mathbf{120}}_{(m=60)} +\overline{\mathbf{132}}_{(m=9)} +\overline{\mathbf{165}}_{(m=25)} +\overline{\mathbf{192}}_{(m=4)} +\overline{\mathbf{210}}_{(m=5)} +\overline{\mathbf{273}}.
\end{align}
\end{widetext}

Flavor and spin : $\begin{tabular}{|c|c|c|c|c|c|}
  \hline
  \quad \quad & \quad \quad & \quad \quad & \quad \quad & \quad \quad & \quad \quad \\
  \hline
  \quad \quad & \quad \quad & \quad \quad & \quad \quad & \quad \quad & \quad \quad \\
  \hline
  \quad \quad & \quad \quad & \quad \quad & \quad \quad & \quad \quad \\
  \cline{1-5}
  \multicolumn{5}{c}{$\overline{\mathbf{3}}(S=\frac{1}{2})$}
\end{tabular}$.\\

\subsection{Diquark case($C=\overline{\mathbf{3}},F=\mathbf{6},S=1$)}
\label{diquark_ASS}

As a second diquark state, we consider color triplet, flavor sextet and spin one diquark.

\subsubsection{One baryon + one diquark}
\label{1b1d2}

The color and flavor $\otimes$ spin states of 5 quarks configuration are same as in Sec.\ref{1b1d}. However, the possible flavor and spin states of multiquark are different after the decomposition. In the case of the four possible diquarks, there are many overlapping parts in relation to decomposition or outer product, so only non-overlapping results are listed here.  \\

Flavor states of 5 quarks :
\begin{align}
  \mathbf{8}\times \mathbf{6}
  =&\overline{\mathbf{3}}+\mathbf{6}+\overline{\mathbf{15}}+\overline{\mathbf{24}}.
\end{align}

Flavor and spin : $\begin{tabular}{|c|c|c}
  \cline{1-2}
  \quad \quad & \quad \quad  \\
  \cline{1-2}
  \quad \quad & \quad \quad \\
  \cline{1-2}
  \quad \quad \\
  \cline{1-1}
  \multicolumn{3}{c}{$\overline{\mathbf{3}}(S=\frac{1}{2},\frac{3}{2})$}
\end{tabular}$,
$\begin{tabular}{|c|c|c|c}
  \cline{1-3}
  \quad \quad & \quad \quad & \quad \quad \\
  \cline{1-3}
  \quad \quad \\
  \cline{1-1}
  \quad \quad \\
  \cline{1-1}
  \multicolumn{4}{c}{$\mathbf{6}(S=\frac{1}{2},\frac{3}{2})$}
\end{tabular}$,
$\begin{tabular}{|c|c|c|c}
  \cline{1-3}
  \quad \quad & \quad \quad & \quad \quad \\
  \cline{1-3}
  \quad \quad & \quad \quad \\
  \cline{1-2}
  \multicolumn{4}{c}{$\overline{\mathbf{15}}(S=\frac{1}{2},\frac{3}{2})$}
\end{tabular}$,
$\begin{tabular}{|c|c|c|c|c}
  \cline{1-4}
  \quad \quad & \quad \quad & \quad \quad & \quad \quad \\
  \cline{1-4}
  \quad \quad \\
  \cline{1-1}
  \multicolumn{5}{c}{$\overline{\mathbf{24}}(S=\frac{1}{2},\frac{3}{2})$}
\end{tabular}$.\\

\subsubsection{Two baryons + one diquark}
\label{2b1d2}

The color and flavor $\otimes$ spin states of 8 quarks configuration are same as in Sec.\ref{2b1d}.\\

\begin{widetext}
Flavor states of 8 quarks :
\begin{align}
  \mathbf{8}\times \mathbf{8} \times \mathbf{6}
  =&\overline{\mathbf{3}}_{(m=3)}+\mathbf{6}_{(m=4)}+\overline{\mathbf{15}}_{(m=5)}+\overline{\mathbf{15'}}+\overline{\mathbf{21}} +\overline{\mathbf{24}}_{(m=4)}
+\overline{\mathbf{42}}_{(m=2)}+\overline{\mathbf{60}}.
\end{align}

Flavor and spin : $\begin{tabular}{|c|c|c|c}
  \cline{1-3}
  \quad \quad & \quad \quad & \quad \quad \\
  \cline{1-3}
  \quad \quad & \quad \quad & \quad \quad \\
  \cline{1-3}
  \quad \quad & \quad \quad \\
  \cline{1-2}
  \multicolumn{4}{c}{$\overline{\mathbf{3}}(S=0,1,2)$}
\end{tabular}$,
$\begin{tabular}{|c|c|c|c|c}
  \cline{1-4}
  \quad \quad & \quad \quad & \quad \quad & \quad \quad  \\
  \cline{1-4}
  \quad \quad & \quad \quad \\
  \cline{1-2}
  \quad \quad & \quad \quad \\
  \cline{1-2}
  \multicolumn{5}{c}{$\mathbf{6}(S=1,2)$}\quad \
\end{tabular}$,
$\begin{tabular}{|c|c|c|c|c}
  \cline{1-4}
  \quad \quad & \quad \quad & \quad \quad & \quad \quad  \\
  \cline{1-4}
  \quad \quad & \quad \quad & \quad \quad \\
  \cline{1-3}
  \quad \quad \\
  \cline{1-1}
  \multicolumn{5}{c}{$\overline{\mathbf{15}}(S=0,1,2)$}
\end{tabular}$,
$\begin{tabular}{|c|c|c|c|}
  \cline{1-4}
  \quad \quad & \quad \quad & \quad \quad & \quad \quad  \\
  \cline{1-4}
  \quad \quad & \quad \quad & \quad \quad & \quad \quad \\
  \cline{1-4}
  \multicolumn{4}{c}{$\overline{\mathbf{15'}}(S=1)$}
\end{tabular}$\ ,
$\begin{tabular}{|c|c|c|c|c|c|}
  \hline
  \quad \quad & \quad \quad & \quad \quad & \quad \quad & \quad \quad & \quad \quad \\
  \hline
  \quad \quad \\
  \cline{1-1}
  \quad \quad \\
  \cline{1-1}
  \multicolumn{6}{c}{$\overline{\mathbf{21}}(S=0)$}\quad \
\end{tabular}$,
$\begin{tabular}{|c|c|c|c|c|c}
  \cline{1-5}
  \quad \quad & \quad \quad & \quad \quad & \quad \quad & \quad \quad \\
  \cline{1-5}
  \quad \quad & \quad \quad \\
  \cline{1-2}
  \quad \quad \\
  \cline{1-1}
  \multicolumn{6}{c}{$\overline{\mathbf{24}}(S=0,1,2)$}\quad \
\end{tabular}$,
$\begin{tabular}{|c|c|c|c|c|c}
  \cline{1-5}
  \quad \quad & \quad \quad & \quad \quad & \quad \quad & \quad \quad \\
  \cline{1-5}
  \quad \quad & \quad \quad & \quad \quad \\
  \cline{1-3}
  \multicolumn{6}{c}{$\overline{\mathbf{42}}(S=0,1,2)$}\quad \
\end{tabular}$,
$\begin{tabular}{|c|c|c|c|c|c|}
  \hline
  \quad \quad & \quad \quad & \quad \quad & \quad \quad & \quad \quad & \quad \quad \\
  \hline
  \quad \quad & \quad \quad \\
  \cline{1-2}
  \multicolumn{6}{c}{$\overline{\mathbf{60}}(S=1)$}
\end{tabular}$.
\end{widetext}

\subsubsection{Three baryons + one diquark}
\label{3b1d2}

The color and flavor $\otimes$ spin states of 11 quarks configuration are same as in Sec.\ref{3b1d}.\\

\begin{widetext}
Flavor states of 11 quarks :
\begin{align}
  \mathbf{8}\times \mathbf{8} \times \mathbf{8} \times \mathbf{6}
  =&\overline{\mathbf{3}}_{(m=12)}+\mathbf{6}_{(m=16)}+\overline{\mathbf{15}}_{(m=24)}+\overline{\mathbf{15'}}_{(m=8)}+\overline{\mathbf{21}}_{(m=6)}+\overline{\mathbf{24}}_{(m=21)}
+\overline{\mathbf{42}}_{(m=15)} +\overline{\mathbf{48}}_{(m=3)} +\overline{\mathbf{60}}_{(m=9)} \nonumber\\
 &+\overline{\mathbf{63}}_{(m=2)} +\overline{\mathbf{90}}_{(m=3)} +\overline{\mathbf{120}}.
\end{align}

Flavor and spin : $\begin{tabular}{|c|c|c|c|c}
  \cline{1-4}
  \quad \quad & \quad \quad & \quad \quad & \quad \quad \\
  \cline{1-4}
  \quad \quad & \quad \quad & \quad \quad & \quad \quad \\
  \cline{1-4}
  \quad \quad & \quad \quad & \quad \quad \\
  \cline{1-3}
  \multicolumn{5}{c}{$\overline{\mathbf{3}}(S=\frac{1}{2},\frac{3}{2},\frac{5}{2})$}
\end{tabular}$,
$\begin{tabular}{|c|c|c|c|c|c}
  \cline{1-5}
  \quad \quad & \quad \quad & \quad \quad & \quad \quad & \quad \quad \\
  \cline{1-5}
  \quad \quad & \quad \quad & \quad \quad \\
  \cline{1-3}
  \quad \quad & \quad \quad & \quad \quad \\
  \cline{1-3}
  \multicolumn{6}{c}{$\mathbf{6}(S=\frac{1}{2},\frac{3}{2},\frac{5}{2})$}
\end{tabular}$,\quad
$\begin{tabular}{|c|c|c|c|c|c}
  \cline{1-5}
  \quad \quad & \quad \quad & \quad \quad & \quad \quad & \quad \quad  \\
  \cline{1-5}
  \quad \quad & \quad \quad & \quad \quad & \quad \quad \\
  \cline{1-4}
  \quad \quad & \quad \quad \\
  \cline{1-2}
  \multicolumn{6}{c}{$\overline{\mathbf{15}}(S=\frac{1}{2},\frac{3}{2},\frac{5}{2})$}
\end{tabular}$,
$\begin{tabular}{|c|c|c|c|c|c}
  \cline{1-5}
  \quad \quad & \quad \quad & \quad \quad & \quad \quad & \quad \quad  \\
  \cline{1-5}
  \quad \quad & \quad \quad & \quad \quad & \quad \quad & \quad \quad \\
  \cline{1-5}
  \quad \quad \\
  \cline{1-1}
  \multicolumn{6}{c}{$\overline{\mathbf{15'}}(S=\frac{1}{2},\frac{3}{2})$}
\end{tabular}$\ , \
$\begin{tabular}{|c|c|c|c|c|c|c}
  \cline{1-6}
  \quad \quad & \quad \quad & \quad \quad & \quad \quad & \quad \quad & \quad \quad \\
  \cline{1-6}
  \quad \quad & \quad \quad & \quad \quad \\
  \cline{1-3}
  \quad \quad & \quad \quad \\
  \cline{1-2}
  \multicolumn{7}{c}{$\overline{\mathbf{24}}(S=\frac{1}{2},\frac{3}{2},\frac{5}{2})$}
\end{tabular}$\quad \quad,\\
$\begin{tabular}{|c|c|c|c|c|c|c}
  \cline{1-6}
  \quad \quad & \quad \quad & \quad \quad & \quad \quad & \quad \quad & \quad \quad \\
  \cline{1-6}
  \quad \quad & \quad \quad & \quad \quad & \quad \quad \\
  \cline{1-4}
  \quad \quad \\
  \cline{1-1}
  \multicolumn{7}{c}{$\overline{\mathbf{42}}(S=\frac{1}{2},\frac{3}{2})$}
\end{tabular}$ \qquad \quad ,
$\begin{tabular}{|c|c|c|c|c|c|c}
  \cline{1-6}
  \quad \quad & \quad \quad & \quad \quad & \quad \quad & \quad \quad & \quad \quad \\
  \cline{1-6}
  \quad \quad & \quad \quad & \quad \quad & \quad \quad & \quad \quad \\
  \cline{1-5}
  \multicolumn{6}{c}{$\overline{\mathbf{48}}(S=\frac{1}{2})$}
\end{tabular}$.
\end{widetext}

\subsubsection{Four baryons + one diquark}
\label{4b1d2}

The color and flavor $\otimes$ spin states of 14 quarks configuration are same as in Sec.\ref{4b1d}.\\

\begin{widetext}
Flavor states of 14 quarks :
\begin{align}
  \mathbf{8}\times \mathbf{8} \times \mathbf{8} \times \mathbf{8} \times \mathbf{6}
  =&\overline{\mathbf{3}}_{(m=52)}+\mathbf{6}_{(m=73)}+\overline{\mathbf{15}}_{(m=120)}+\overline{\mathbf{15'}}_{(m=50)}+\overline{\mathbf{21}}_{(m=38)}+\overline{\mathbf{24}}_{(m=112)}
+\overline{\mathbf{36}}_{(m=3)}+\overline{\mathbf{42}}_{(m=98)} \nonumber\\
&+\overline{\mathbf{45}}_{(m=2)} +\overline{\mathbf{48}}_{(m=32)}+\overline{\mathbf{60}}_{(m=66)}+\overline{\mathbf{63}}_{(m=20)}+\overline{\mathbf{90}}_{(m=34)} + +\overline{\mathbf{105}}_{(m=6)} +\overline{\mathbf{120}}_{(m=16)} \nonumber\\
& +\overline{\mathbf{132}}_{(m=3)} +\overline{\mathbf{165}}_{(m=4)} +\overline{\mathbf{210}}.
\end{align}

Flavor and spin : $\begin{tabular}{|c|c|c|c|c|c}
  \hline
  \quad \quad & \quad \quad & \quad \quad & \quad \quad & \quad \quad \\
  \hline
  \quad \quad & \quad \quad & \quad \quad & \quad \quad & \quad \quad \\
  \hline
  \quad \quad & \quad \quad & \quad \quad & \quad \quad\\
  \cline{1-4}
  \multicolumn{5}{c}{$\overline{\mathbf{3}}(S=0,1)$}
\end{tabular}$,
$\begin{tabular}{|c|c|c|c|c|c|}
  \hline
  \quad \quad & \quad \quad & \quad \quad & \quad \quad & \quad \quad & \quad \quad \\
  \hline
  \quad \quad & \quad \quad & \quad \quad & \quad \quad \\
  \cline{1-4}
  \quad \quad & \quad \quad & \quad \quad & \quad \quad \\
  \cline{1-4}
  \multicolumn{5}{c}{$\mathbf{6}(S=1)$}
\end{tabular}$,
$\begin{tabular}{|c|c|c|c|c|c|}
  \hline
  \quad \quad & \quad \quad & \quad \quad & \quad \quad & \quad \quad & \quad \quad\\
  \hline
  \quad \quad & \quad \quad & \quad \quad & \quad \quad & \quad \quad \\
  \cline{1-5}
  \quad \quad & \quad \quad & \quad \quad \\
  \cline{1-3}
  \multicolumn{6}{c}{$\overline{\mathbf{15}}(S=0,1,2)$}
\end{tabular}$,
$\begin{tabular}{|c|c|c|c|c|c|}
  \hline
  \quad \quad & \quad \quad & \quad \quad & \quad \quad & \quad \quad & \quad \quad \\
  \hline
  \quad \quad & \quad \quad & \quad \quad & \quad \quad & \quad \quad & \quad \quad \\
  \hline
  \quad \quad & \quad \quad \\
  \cline{1-2}
  \multicolumn{6}{c}{$\overline{\mathbf{15'}}(S=1)$}
\end{tabular}$.
\end{widetext}

\subsubsection{Five baryons + one diquark}
\label{5b1d2}

The color and flavor $\otimes$ spin states of 17 quarks configuration are same as in Sec.\ref{5b1d}.\\

\begin{widetext}
Flavor states of 17 quarks :
\begin{align}
  \mathbf{8}\times \mathbf{8}\times \mathbf{8} \times \mathbf{8} \times \mathbf{8} \times \mathbf{6}
  =& \overline{\mathbf{3}}_{(m=245)}+\mathbf{6}_{(m=357)}+\overline{\mathbf{15}}_{(m=625)}+\overline{\mathbf{15'}}_{(m=300)}+ \overline{\mathbf{21}}_{(m=236)}+\overline{\mathbf{24}}_{(m=619)}
+ \overline{\mathbf{36}}_{(m=41)} \nonumber\\
& +\overline{\mathbf{42}}_{(m=610)} + \overline{\mathbf{45}}_{(m=25)}+ \overline{\mathbf{48}}_{(m=255)} +\overline{\mathbf{60}}_{(m=450)} +\overline{\mathbf{63}}_{(m=165)} +\overline{\mathbf{90}}_{(m=290)} +\overline{\mathbf{99}}_{(m=9)}  \nonumber\\
& +\overline{\mathbf{105}}_{(m=85)} +\overline{\mathbf{120}}_{(m=160)} +\overline{\mathbf{120'}}_{(m=5)} +\overline{\mathbf{132}}_{(m=45)} +\overline{\mathbf{165}}_{(m=65)} +\overline{\mathbf{192}}_{(m=10)} +\overline{\mathbf{210}}_{(m=25)} \nonumber\\
& +\overline{\mathbf{234}}_{(m=4)} +\overline{\mathbf{273}}_{(m=5)} +\overline{\mathbf{336}}.
\end{align}
\end{widetext}

Flavor and spin : $\begin{tabular}{|c|c|c|c|c|c|}
  \hline
  \quad \quad & \quad \quad & \quad \quad & \quad \quad & \quad \quad & \quad \quad \\
  \hline
  \quad \quad & \quad \quad & \quad \quad & \quad \quad & \quad \quad & \quad \quad \\
  \hline
  \quad \quad & \quad \quad & \quad \quad & \quad \quad & \quad \quad \\
  \cline{1-5}
  \multicolumn{5}{c}{$\overline{\mathbf{3}}(S=\frac{1}{2})$}
\end{tabular}$.

\subsection{Diquark case($C=\mathbf{6},F=\overline{\mathbf{3}},S=1$)}
\label{diquark_SAS}

As a third diquark state, we consider the color sextet, flavor antitriplet and spin one diquark.

\subsubsection{One baryon + one diquark}
\label{1b1d3}

The color and flavor $\otimes$ spin states of 5 quarks configuration are as follows.\\

Color : $\begin{tabular}{|c|c|c|}
  \hline
  \quad \quad & \quad \quad & \quad \quad \\
  \hline
  \quad \quad \\
  \cline{1-1}
  \quad \quad \\
  \cline{1-1}
\end{tabular}$, \quad
Flavor $\otimes$ Spin : $\begin{tabular}{|c|c|c|}
  \hline
  \quad \quad & \quad \quad & \quad \quad \\
  \hline
  \quad \quad \\
  \cline{1-1}
  \quad \quad \\
  \cline{1-1}
\end{tabular}.$\\
The flavor $\otimes$ spin coupling state [3,1,1] with SU(6) can be decomposed into the states with the flavor SU(3) and the spin SU(2) as follows.
\begin{widetext}
\begin{align}
[3,1,1]_{FS} &=[4,1]_F \otimes [4,1]_S + [4,1]_F \otimes [3,2]_S + [3,2]_F \otimes [4,1]_S + [3,2]_F \otimes [3,2]_S + [3,1,1]_F \otimes [5]_S + [3,1,1]_F \otimes [4,1]_S  \nonumber\\
  &\quad + [3,1,1]_F \otimes [3,2]_{S(m=2)} + [2,2,1]_F \otimes [4,1]_S + [2,2,1]_F \otimes [3,2]_S.
\end{align}

Flavor and spin : $\begin{tabular}{|c|c|c}
  \cline{1-2}
  \quad \quad & \quad \quad  \\
  \cline{1-2}
  \quad \quad & \quad \quad \\
  \cline{1-2}
  \quad \quad \\
  \cline{1-1}
  \multicolumn{3}{c}{$\overline{\mathbf{3}}(S=\frac{1}{2},\frac{3}{2})$}
\end{tabular}$,
$\begin{tabular}{|c|c|c|c}
  \cline{1-3}
  \quad \quad & \quad \quad & \quad \quad \\
  \cline{1-3}
  \quad \quad \\
  \cline{1-1}
  \quad \quad \\
  \cline{1-1}
  \multicolumn{4}{c}{$\mathbf{6}(S=\frac{1}{2},\frac{3}{2})$}
\end{tabular}$,
$\begin{tabular}{|c|c|c|c}
  \cline{1-3}
  \quad \quad & \quad \quad & \quad \quad \\
  \cline{1-3}
  \quad \quad & \quad \quad \\
  \cline{1-2}
  \multicolumn{4}{c}{$\overline{\mathbf{15}}(S=\frac{1}{2},\frac{3}{2})$}
\end{tabular}$.
\end{widetext}

\subsubsection{Two baryons + one diquark}
\label{2b1d3}

The color and flavor $\otimes$ spin states of 8 quarks configuration are as follows.\\

Color : $\begin{tabular}{|c|c|c|c|}
  \hline
  \quad \quad & \quad \quad & \quad \quad & \quad \quad \\
  \hline
  \quad \quad & \quad \quad \\
  \cline{1-2}
  \quad \quad & \quad \quad \\
  \cline{1-2}
\end{tabular}$, \quad
Flavor $\otimes$ Spin : $\begin{tabular}{|c|c|c|}
  \hline
  \quad \quad & \quad \quad & \quad \quad \\
  \hline
  \quad \quad & \quad \quad & \quad \quad \\
  \hline
  \quad \quad \\
  \cline{1-1}
  \quad \quad \\
  \cline{1-1}
\end{tabular}.$\\
The flavor $\otimes$ spin coupling state [3,3,1,1] with SU(6) can be decomposed into the states with the flavor SU(3) and the spin SU(2) as follows.
\begin{widetext}
\begin{align}
  [3,3,1,1]_{FS} &= [6,2]_F \otimes [4,4]_S + [6,1,1]_F \otimes [5,3]_S + [5,3]_F \otimes [5,3]_S + [5,2,1]_F \otimes [6,2]_S + [5,2,1]_F \otimes [5,3]_{S(m=2)}  \nonumber\\
   & \quad + [5,2,1]_F \otimes [4,4]_S + [4,4]_F \otimes [6,2]_S + [4,4]_F \otimes [4,4]_S + [4,3,1]_F \otimes [7,1]_S + [4,3,1]_F \otimes [6,2]_{S(m=2)}  \nonumber\\
    &\quad + [4,3,1]_F \otimes [5,3]_{S(m=3)} + [4,3,1]_F \otimes [4,4]_S + [4,2,2]_F \otimes [6,2]_{S(m=2)} + [4,2,2]_F \otimes [5,3]_S  \nonumber\\
    &\quad + [4,2,2]_F \otimes [4,4]_{S(m=2)} + [3,3,2]_F \otimes [7,1]_S + [3,3,2]_F \otimes [6,2]_S + [3,3,2]_F \otimes [5,3]_{S(m=2)}.
\end{align}

Flavor and spin : $\begin{tabular}{|c|c|c|c}
  \cline{1-3}
  \quad \quad & \quad \quad & \quad \quad \\
  \cline{1-3}
  \quad \quad & \quad \quad & \quad \quad \\
  \cline{1-3}
  \quad \quad & \quad \quad \\
  \cline{1-2}
  \multicolumn{4}{c}{$\overline{\mathbf{3}}(S=1,2)$}
\end{tabular}$,
$\begin{tabular}{|c|c|c|c|c}
  \cline{1-4}
  \quad \quad & \quad \quad & \quad \quad & \quad \quad  \\
  \cline{1-4}
  \quad \quad & \quad \quad \\
  \cline{1-2}
  \quad \quad & \quad \quad \\
  \cline{1-2}
  \multicolumn{5}{c}{$\mathbf{6}(S=0,1,2)$} \
\end{tabular}$,
$\begin{tabular}{|c|c|c|c|c}
  \cline{1-4}
  \quad \quad & \quad \quad & \quad \quad & \quad \quad  \\
  \cline{1-4}
  \quad \quad & \quad \quad & \quad \quad \\
  \cline{1-3}
  \quad \quad \\
  \cline{1-1}
  \multicolumn{5}{c}{$\overline{\mathbf{15}}(S=0,1,2)$}
\end{tabular}$,
$\begin{tabular}{|c|c|c|c|}
  \cline{1-4}
  \quad \quad & \quad \quad & \quad \quad & \quad \quad  \\
  \cline{1-4}
  \quad \quad & \quad \quad & \quad \quad & \quad \quad \\
  \cline{1-4}
  \multicolumn{4}{c}{$\overline{\mathbf{15'}}(S=0,2)$}
\end{tabular}$\ ,
$\begin{tabular}{|c|c|c|c|c|c}
  \cline{1-5}
  \quad \quad & \quad \quad & \quad \quad & \quad \quad & \quad \quad \\
  \cline{1-5}
  \quad \quad & \quad \quad \\
  \cline{1-2}
  \quad \quad \\
  \cline{1-1}
  \multicolumn{6}{c}{$\overline{\mathbf{24}}(S=0,1,2)$}\quad \
\end{tabular}$,\\
$\begin{tabular}{|c|c|c|c|c|c}
  \cline{1-5}
  \quad \quad & \quad \quad & \quad \quad & \quad \quad & \quad \quad \\
  \cline{1-5}
  \quad \quad & \quad \quad & \quad \quad \\
  \cline{1-3}
  \multicolumn{6}{c}{$\overline{\mathbf{42}}(S=1)$}
\end{tabular}$.
\end{widetext}

\subsubsection{Three baryons + one diquark}
\label{3b1d3}

The color and flavor $\otimes$ spin states of 11 quarks configuration are as follows.\\

Color : $\begin{tabular}{|c|c|c|c|c|}
  \hline
  \quad \quad & \quad \quad & \quad \quad & \quad \quad & \quad \quad \\
  \hline
  \quad \quad & \quad \quad & \quad \quad \\
  \cline{1-3}
  \quad \quad & \quad \quad & \quad \quad \\
  \cline{1-3}
\end{tabular}$, \quad
Flavor $\otimes$ Spin : $\begin{tabular}{|c|c|c|}
  \hline
  \quad \quad & \quad \quad & \quad \quad  \\
  \hline
  \quad \quad & \quad \quad & \quad \quad \\
  \hline
  \quad \quad & \quad \quad & \quad \quad \\
  \hline
  \quad \quad \\
  \cline{1-1}
  \quad \quad \\
  \cline{1-1}
\end{tabular}$\\
The flavor $\otimes$ spin coupling state [3,3,3,1,1] with SU(6) can be decomposed into the states with the flavor SU(3) and the spin SU(2) as follows.
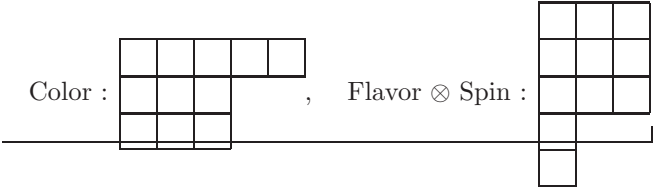
\begin{widetext}
\begin{align}
  [3,3,3,1,1]_{FS} &= [6,4,1]_F \otimes [7,4]_S + [6,4,1]_F \otimes [6,5]_S + [6,3,2]_F \otimes [7,4]_S + [6,3,2]_F \otimes [6,5]_S + [5,5,1]_F \otimes [6,5]_S  \nonumber\\
   & \quad + [5,4,2]_F \otimes [8,3]_S  + [5,4,2]_F \otimes [7,4]_{S(m=2)} + [5,4,2]_F \otimes [6,5]_{S(m=2)} + [5,3,3]_F \otimes [8,3]_S  \nonumber\\
    &\quad + [5,3,3]_F \otimes [7,4]_S + [5,3,3]_F \otimes [6,5]_S + [4,4,3]_F \otimes [9,2]_S + [4,4,3]_F \otimes [8,3]_S  \nonumber\\
    &\quad + [4,4,3]_F \otimes [7,4]_{S(m=2)} + [4,4,3]_F \otimes [6,5]_S.
\end{align}

Flavor and spin : $\begin{tabular}{|c|c|c|c|c}
  \cline{1-4}
  \quad \quad & \quad \quad & \quad \quad & \quad \quad \\
  \cline{1-4}
  \quad \quad & \quad \quad & \quad \quad & \quad \quad \\
  \cline{1-4}
  \quad \quad & \quad \quad & \quad \quad \\
  \cline{1-3}
  \multicolumn{5}{c}{$\overline{\mathbf{3}}(S=\frac{1}{2},\frac{3}{2},\frac{5}{2})$}
\end{tabular}$,
$\begin{tabular}{|c|c|c|c|c|c}
  \cline{1-5}
  \quad \quad & \quad \quad & \quad \quad & \quad \quad & \quad \quad \\
  \cline{1-5}
  \quad \quad & \quad \quad & \quad \quad \\
  \cline{1-3}
  \quad \quad & \quad \quad & \quad \quad \\
  \cline{1-3}
  \multicolumn{6}{c}{$\mathbf{6}(S=\frac{1}{2},\frac{3}{2},\frac{5}{2})$}
\end{tabular}$\quad ,\quad
$\begin{tabular}{|c|c|c|c|c|c}
  \cline{1-5}
  \quad \quad & \quad \quad & \quad \quad & \quad \quad & \quad \quad  \\
  \cline{1-5}
  \quad \quad & \quad \quad & \quad \quad & \quad \quad \\
  \cline{1-4}
  \quad \quad & \quad \quad \\
  \cline{1-2}
  \multicolumn{6}{c}{$\overline{\mathbf{15}}(S=\frac{1}{2},\frac{3}{2},\frac{5}{2})$}
\end{tabular}$\quad,
$\begin{tabular}{|c|c|c|c|c|c}
  \cline{1-5}
  \quad \quad & \quad \quad & \quad \quad & \quad \quad & \quad \quad  \\
  \cline{1-5}
  \quad \quad & \quad \quad & \quad \quad & \quad \quad & \quad \quad \\
  \cline{1-5}
  \quad \quad \\
  \cline{1-1}
  \multicolumn{6}{c}{$\overline{\mathbf{15'}}(S=\frac{1}{2})$}
\end{tabular}$\quad , \quad
$\begin{tabular}{|c|c|c|c|c|c|c}
  \cline{1-6}
  \quad \quad & \quad \quad & \quad \quad & \quad \quad & \quad \quad & \quad \quad \\
  \cline{1-6}
  \quad \quad & \quad \quad & \quad \quad \\
  \cline{1-3}
  \quad \quad & \quad \quad \\
  \cline{1-2}
  \multicolumn{7}{c}{$\overline{\mathbf{24}}(S=\frac{1}{2},\frac{3}{2})$}
\end{tabular}$\quad \quad,\\
$\begin{tabular}{|c|c|c|c|c|c|c}
  \cline{1-6}
  \quad \quad & \quad \quad & \quad \quad & \quad \quad & \quad \quad & \quad \quad \\
  \cline{1-6}
  \quad \quad & \quad \quad & \quad \quad & \quad \quad \\
  \cline{1-4}
  \quad \quad \\
  \cline{1-1}
  \multicolumn{7}{c}{$\overline{\mathbf{42}}(S=\frac{1}{2},\frac{3}{2})$}
\end{tabular}$.
\end{widetext}

\subsubsection{Four baryons + one diquark}
\label{4b1d3}

The color and flavor $\otimes$ spin states of 14 quarks configuration are as follows.\\

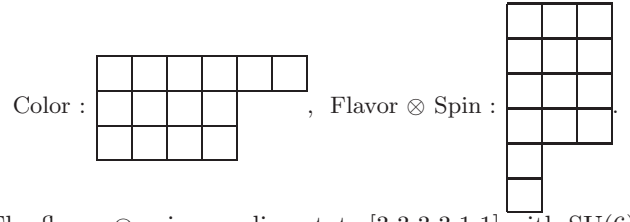
\begin{small}
Color : $\begin{tabular}{|c|c|c|c|c|c|}
  \hline
  \quad \quad & \quad \quad & \quad \quad & \quad \quad & \quad \quad & \quad \quad \\
  \hline
  \quad \quad & \quad \quad & \quad \quad & \quad \quad \\
  \cline{1-4}
  \quad \quad & \quad \quad & \quad \quad & \quad \quad \\
  \cline{1-4}
\end{tabular}$, \
Flavor $\otimes$ Spin : $\begin{tabular}{|c|c|c|}
  \hline
  \quad \quad & \quad \quad & \quad \quad  \\
  \hline
  \quad \quad & \quad \quad & \quad \quad  \\
  \hline
  \quad \quad & \quad \quad & \quad \quad  \\
  \hline
  \quad \quad & \quad \quad & \quad \quad  \\
  \hline
  \quad \quad \\
  \cline{1-1}
  \quad \quad \\
  \cline{1-1}
\end{tabular}.$
\end{small}\\
The flavor $\otimes$ spin coupling state [3,3,3,3,1,1] with SU(6) can be decomposed into the states with the flavor SU(3) and the spin SU(2) as follows.
\begin{widetext}
\begin{align}
[3,3,3,3,1,1]_{FS} &=[6,6,2]_F \otimes [7,7]_S + [6,5,3]_F \otimes [8,6]_S + [6,4,4]_F \otimes [9,5]_S + [6,4,4]_F \otimes [7,7]_S + [5,5,4]_F \otimes [8,6]_S.
\end{align}

Flavor and spin : $\begin{tabular}{|c|c|c|c|c|c}
  \hline
  \quad \quad & \quad \quad & \quad \quad & \quad \quad & \quad \quad \\
  \hline
  \quad \quad & \quad \quad & \quad \quad & \quad \quad & \quad \quad \\
  \hline
  \quad \quad & \quad \quad & \quad \quad & \quad \quad\\
  \cline{1-4}
  \multicolumn{5}{c}{$\overline{\mathbf{3}}(S=1)$}
\end{tabular}$,
$\begin{tabular}{|c|c|c|c|c|c|}
  \hline
  \quad \quad & \quad \quad & \quad \quad & \quad \quad & \quad \quad & \quad \quad \\
  \hline
  \quad \quad & \quad \quad & \quad \quad & \quad \quad \\
  \cline{1-4}
  \quad \quad & \quad \quad & \quad \quad & \quad \quad \\
  \cline{1-4}
  \multicolumn{5}{c}{$\mathbf{6}(S=0,2)$}
\end{tabular}$,
$\begin{tabular}{|c|c|c|c|c|c|}
  \hline
  \quad \quad & \quad \quad & \quad \quad & \quad \quad & \quad \quad & \quad \quad\\
  \hline
  \quad \quad & \quad \quad & \quad \quad & \quad \quad & \quad \quad \\
  \cline{1-5}
  \quad \quad & \quad \quad & \quad \quad \\
  \cline{1-3}
  \multicolumn{6}{c}{$\overline{\mathbf{15}}(S=1)$}
\end{tabular}$,
$\begin{tabular}{|c|c|c|c|c|c|}
  \hline
  \quad \quad & \quad \quad & \quad \quad & \quad \quad & \quad \quad & \quad \quad \\
  \hline
  \quad \quad & \quad \quad & \quad \quad & \quad \quad & \quad \quad & \quad \quad \\
  \hline
  \quad \quad & \quad \quad \\
  \cline{1-2}
  \multicolumn{6}{c}{$\overline{\mathbf{15'}}(S=0)$}
\end{tabular}$.
\end{widetext}

\subsubsection{Five baryons + one diquark}
\label{5b1d3}

The color and flavor $\otimes$ spin states of 17 quarks configuration are as follows.\\
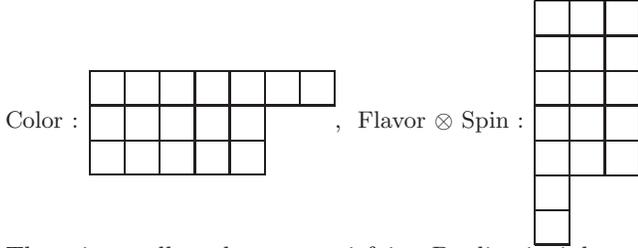
\begin{small}
Color : $\begin{tabular}{|c|c|c|c|c|c|c|}
  \hline
  \quad \quad & \quad \quad & \quad \quad & \quad \quad & \quad \quad & \quad \quad & \quad \quad \\
  \hline
  \quad \quad & \quad \quad & \quad \quad & \quad \quad & \quad \quad \\
  \cline{1-5}
  \quad \quad & \quad \quad & \quad \quad & \quad \quad & \quad \quad \\
  \cline{1-5}
\end{tabular}$, \
Flavor $\otimes$ Spin : $\begin{tabular}{|c|c|c|}
  \hline
  \quad \quad & \quad \quad & \quad \quad  \\
  \hline
  \quad \quad & \quad \quad & \quad \quad  \\
  \hline
  \quad \quad & \quad \quad & \quad \quad  \\
  \hline
  \quad \quad & \quad \quad & \quad \quad  \\
  \hline
  \quad \quad & \quad \quad & \quad \quad  \\
  \hline
  \quad \quad \\
  \cline{1-1}
  \quad \quad \\
  \cline{1-1}
\end{tabular}.$\\
\end{small}
There is no allowed states satisfying Pauli principle.

\subsection{Diquark case($C=\mathbf{6},F=\mathbf{6},S=0$)}
\label{diquark_SSA}
As a final diquark case, we consider color sextet, flavor sextet and spin zero diquark.

\subsubsection{One baryon + one diquark}
\label{1b1d4}

The color and flavor $\otimes$ spin states of 5 quarks configuration are same as in Sec.\ref{1b1d3}. In addition, flavor states of 5 quarks are same as in Sec.\ref{1b1d2}. Therefore, for the rest of the cases, we only represent the possible flavor and spin states.\\

Flavor and spin : $\begin{tabular}{|c|c|c}
  \cline{1-2}
  \quad \quad & \quad \quad  \\
  \cline{1-2}
  \quad \quad & \quad \quad \\
  \cline{1-2}
  \quad \quad \\
  \cline{1-1}
  \multicolumn{3}{c}{$\overline{\mathbf{3}}(S=\frac{1}{2})$}
\end{tabular}$,
$\begin{tabular}{|c|c|c|c}
  \cline{1-3}
  \quad \quad & \quad \quad & \quad \quad \\
  \cline{1-3}
  \quad \quad \\
  \cline{1-1}
  \quad \quad \\
  \cline{1-1}
  \multicolumn{4}{c}{$\mathbf{6}(S=\frac{1}{2})$}
\end{tabular}$,
$\begin{tabular}{|c|c|c|c}
  \cline{1-3}
  \quad \quad & \quad \quad & \quad \quad \\
  \cline{1-3}
  \quad \quad & \quad \quad \\
  \cline{1-2}
  \multicolumn{4}{c}{$\overline{\mathbf{15}}(S=\frac{1}{2})$}
\end{tabular}$,
$\begin{tabular}{|c|c|c|c|c}
  \cline{1-4}
  \quad \quad & \quad \quad & \quad \quad & \quad \quad \\
  \cline{1-4}
  \quad \quad \\
  \cline{1-1}
  \multicolumn{5}{c}{$\overline{\mathbf{24}}(S=\frac{1}{2})$}
\end{tabular}$.

\subsubsection{Two baryons + one diquark}
\label{2b1d4}

Flavor and spin : $\begin{tabular}{|c|c|c|c}
  \cline{1-3}
  \quad \quad & \quad \quad & \quad \quad \\
  \cline{1-3}
  \quad \quad & \quad \quad & \quad \quad \\
  \cline{1-3}
  \quad \quad & \quad \quad \\
  \cline{1-2}
  \multicolumn{4}{c}{$\overline{\mathbf{3}}(S=1)$}
\end{tabular}$,
$\begin{tabular}{|c|c|c|c|c}
  \cline{1-4}
  \quad \quad & \quad \quad & \quad \quad & \quad \quad  \\
  \cline{1-4}
  \quad \quad & \quad \quad \\
  \cline{1-2}
  \quad \quad & \quad \quad \\
  \cline{1-2}
  \multicolumn{5}{c}{$\mathbf{6}(S=0,1)$}
\end{tabular}$,
$\begin{tabular}{|c|c|c|c|c}
  \cline{1-4}
  \quad \quad & \quad \quad & \quad \quad & \quad \quad  \\
  \cline{1-4}
  \quad \quad & \quad \quad & \quad \quad \\
  \cline{1-3}
  \quad \quad \\
  \cline{1-1}
  \multicolumn{5}{c}{$\overline{\mathbf{15}}(S=0,1)$}
\end{tabular}$,
$\begin{tabular}{|c|c|c|c|c}
  \cline{1-4}
  \quad \quad & \quad \quad & \quad \quad & \quad \quad  \\
  \cline{1-4}
  \quad \quad & \quad \quad & \quad \quad & \quad \quad \\
  \cline{1-4}
  \multicolumn{5}{c}{$\overline{\mathbf{15'}}(S=0)$}
\end{tabular}$\ ,
$\begin{tabular}{|c|c|c|c|c|c|}
  \hline
  \quad \quad & \quad \quad & \quad \quad & \quad \quad & \quad \quad & \quad \quad \\
  \hline
  \quad \quad \\
  \cline{1-1}
  \quad \quad \\
  \cline{1-1}
  \multicolumn{6}{c}{$\overline{\mathbf{21}}(S=1)$}\quad \
\end{tabular}$,
$\begin{tabular}{|c|c|c|c|c|c}
  \cline{1-5}
  \quad \quad & \quad \quad & \quad \quad & \quad \quad & \quad \quad \\
  \cline{1-5}
  \quad \quad & \quad \quad \\
  \cline{1-2}
  \quad \quad \\
  \cline{1-1}
  \multicolumn{6}{c}{$\overline{\mathbf{24}}(S=0,1)$}\quad \
\end{tabular}$,
$\begin{tabular}{|c|c|c|c|c|c}
  \cline{1-5}
  \quad \quad & \quad \quad & \quad \quad & \quad \quad & \quad \quad \\
  \cline{1-5}
  \quad \quad & \quad \quad & \quad \quad \\
  \cline{1-3}
  \multicolumn{6}{c}{$\overline{\mathbf{42}}(S=1)$}\quad \quad
\end{tabular}$,
$\begin{tabular}{|c|c|c|c|c|c|}
  \hline
  \quad \quad & \quad \quad & \quad \quad & \quad \quad & \quad \quad & \quad \quad \\
  \hline
  \quad \quad & \quad \quad \\
  \cline{1-2}
  \multicolumn{6}{c}{$\overline{\mathbf{60}}(S=0)$}
\end{tabular}$.

\subsubsection{Three baryons + one diquark}
\label{3b1d4}

Flavor and spin : $\begin{tabular}{|c|c|c|c|c}
  \cline{1-4}
  \quad \quad & \quad \quad & \quad \quad & \quad \quad \\
  \cline{1-4}
  \quad \quad & \quad \quad & \quad \quad & \quad \quad \\
  \cline{1-4}
  \quad \quad & \quad \quad & \quad \quad \\
  \cline{1-3}
  \multicolumn{5}{c}{$\overline{\mathbf{3}}(S=\frac{1}{2},\frac{3}{2})$}
\end{tabular}$,
$\begin{tabular}{|c|c|c|c|c|}
  \hline
  \quad \quad & \quad \quad & \quad \quad & \quad \quad & \quad \quad \\
  \hline
  \quad \quad & \quad \quad & \quad \quad \\
  \cline{1-3}
  \quad \quad & \quad \quad & \quad \quad \\
  \cline{1-3}
  \multicolumn{5}{c}{$\mathbf{6}(S=\frac{1}{2},\frac{3}{2})$}
\end{tabular}$,
$\begin{tabular}{|c|c|c|c|c|}
  \hline
  \quad \quad & \quad \quad & \quad \quad & \quad \quad & \quad \quad  \\
  \hline
  \quad \quad & \quad \quad & \quad \quad & \quad \quad \\
  \cline{1-4}
  \quad \quad & \quad \quad \\
  \cline{1-2}
  \multicolumn{5}{c}{$\overline{\mathbf{15}}(S=\frac{1}{2},\frac{3}{2})$}
\end{tabular}$,
$\begin{tabular}{|c|c|c|c|c|}
  \hline
  \quad \quad & \quad \quad & \quad \quad & \quad \quad & \quad \quad  \\
  \hline
  \quad \quad & \quad \quad & \quad \quad & \quad \quad & \quad \quad \\
  \cline{1-5}
  \quad \quad \\
  \cline{1-1}
  \multicolumn{5}{c}{$\overline{\mathbf{15'}}(S=\frac{1}{2})$}
\end{tabular}$,
$\begin{tabular}{|c|c|c|c|c|c|}
  \hline
  \quad \quad & \quad \quad & \quad \quad & \quad \quad & \quad \quad & \quad \quad \\
  \hline
  \quad \quad & \quad \quad & \quad \quad \\
  \cline{1-3}
  \quad \quad & \quad \quad \\
  \cline{1-2}
  \multicolumn{6}{c}{$\overline{\mathbf{24}}(S=\frac{1}{2},\frac{3}{2})$}
\end{tabular}$,\\
$\begin{tabular}{|c|c|c|c|c|c|}
  \hline
  \quad \quad & \quad \quad & \quad \quad & \quad \quad & \quad \quad & \quad \quad \\
  \hline
  \quad \quad & \quad \quad & \quad \quad & \quad \quad \\
  \cline{1-4}
  \quad \quad \\
  \cline{1-1}
  \multicolumn{6}{c}{$\overline{\mathbf{42}}(S=\frac{1}{2},\frac{3}{2})$}
\end{tabular}$.

\subsubsection{Four baryons + one diquark}
\label{4b1d4}

\begin{small}
Flavor and spin : $\begin{tabular}{|c|c|c|c|c|c}
  \hline
  \quad \quad & \quad \quad & \quad \quad & \quad \quad & \quad \quad \\
  \hline
  \quad \quad & \quad \quad & \quad \quad & \quad \quad & \quad \quad \\
  \hline
  \quad \quad & \quad \quad & \quad \quad & \quad \quad\\
  \cline{1-4}
  \multicolumn{5}{c}{$\overline{\mathbf{3}}(S=1)$}
\end{tabular}$,
$\begin{tabular}{|c|c|c|c|c|c|}
  \hline
  \quad \quad & \quad \quad & \quad \quad & \quad \quad & \quad \quad & \quad \quad \\
  \hline
  \quad \quad & \quad \quad & \quad \quad & \quad \quad \\
  \cline{1-4}
  \quad \quad & \quad \quad & \quad \quad & \quad \quad \\
  \cline{1-4}
  \multicolumn{5}{c}{$\mathbf{6}(S=0,2)$}
\end{tabular}$,
$\begin{tabular}{|c|c|c|c|c|c|}
  \hline
  \quad \quad & \quad \quad & \quad \quad & \quad \quad & \quad \quad & \quad \quad\\
  \hline
  \quad \quad & \quad \quad & \quad \quad & \quad \quad & \quad \quad \\
  \cline{1-5}
  \quad \quad & \quad \quad & \quad \quad \\
  \cline{1-3}
  \multicolumn{6}{c}{$\overline{\mathbf{15}}(S=1)$}
\end{tabular}$,
$\begin{tabular}{|c|c|c|c|c|c|}
  \hline
  \quad \quad & \quad \quad & \quad \quad & \quad \quad & \quad \quad & \quad \quad \\
  \hline
  \quad \quad & \quad \quad & \quad \quad & \quad \quad & \quad \quad & \quad \quad \\
  \hline
  \quad \quad & \quad \quad \\
  \cline{1-2}
  \multicolumn{6}{c}{$\overline{\mathbf{15'}}(S=0)$}
\end{tabular}$.
\end{small}

\subsubsection{Five baryons + one diquark}
\label{5b1d4}

There is no allowed states satisfying Pauli principle.

\subsection{Three correlated diquarks}

When we consider the interaction involving a free quark or a free diquark, the problem of the infinite thermal Wilson line in the confining phase arises. To avoid that problem, we consider the color singlet state of three diquarks additionally. Among several possible states, the most attractive one is flavor singlet and spin zero state which is the same quantum number as H-dibaryon. In this study, we construct the three correlated diquarks using most stable diquark state which is color antitriplet, flavor antitriplet and spin zero. Also, it should be noted that this state is not allowed in flavor SU(2) case because it should contain two strange quarks. 

\subsubsection{One baryon + three diquarks}

The color and flavor $\otimes$ spin states of 9 quarks composed of one baryon and three correlated diquarks are as follows.\\
\\
Color : $\begin{tabular}{|c|c|c|}
  \hline
  \quad \quad & \quad \quad & \quad \quad \\
  \hline
  \quad \quad & \quad \quad & \quad \quad \\
  \hline
  \quad \quad & \quad \quad & \quad \quad \\
  \hline
\end{tabular}$, \quad
Flavor $\otimes$ Spin : $\begin{tabular}{|c|c|c|}
  \hline
  \quad \quad & \quad \quad & \quad \quad  \\
  \hline
  \quad \quad & \quad \quad & \quad \quad  \\
  \hline
  \quad \quad & \quad \quad & \quad \quad  \\
  \hline
\end{tabular}.$\\

Flavor states of 9 quarks :
\begin{align}
\mathbf{8}\times \mathbf{1} = \mathbf{8}.
\end{align}

Using the decomposition in octet baryon case, we can find that there is one possible flavor and spin states as follows.\\

Flavor and spin :
$\begin{tabular}{|c|c|c|c|}
  \hline
  \quad \quad & \quad \quad & \quad \quad & \quad \quad \\
  \hline
  \quad \quad & \quad \quad & \quad \quad \\
  \cline{1-3}
  \quad \quad & \quad \quad \\
  \cline{1-2}
  \multicolumn{4}{c}{$\mathbf{8}(S=\frac{1}{2})$}
\end{tabular}$.

\subsubsection{Two baryons + three diquarks}

The color and flavor $\otimes$ spin states of 12 quarks are as follows.\\
\\
Color : $\begin{tabular}{|c|c|c|c|}
  \hline
  \quad \quad & \quad \quad & \quad \quad & \quad \quad \\
  \hline
  \quad \quad & \quad \quad & \quad \quad & \quad \quad \\
  \hline
  \quad \quad & \quad \quad & \quad \quad & \quad \quad \\
  \hline
\end{tabular}$, \quad
Flavor $\otimes$ Spin : $\begin{tabular}{|c|c|c|}
  \hline
  \quad \quad & \quad \quad & \quad \quad  \\
  \hline
  \quad \quad & \quad \quad & \quad \quad  \\
  \hline
  \quad \quad & \quad \quad & \quad \quad  \\
  \hline
  \quad \quad & \quad \quad & \quad \quad  \\
  \hline
\end{tabular}.$\\

Flavor states of 12 quarks :
\begin{align}
\mathbf{8}\times \mathbf{8} \times \mathbf{1} = \mathbf{1}+\mathbf{8}_{(m=2)}+\mathbf{10}+\overline{\mathbf{10}}+\mathbf{27}.
\end{align}

Using the decomposition in octet baryon case, we can determine the possible flavor and spin states as follows.\\

\begin{small}
Flavor and spin :
$\begin{tabular}{|c|c|c|c|}
  \cline{1-4}
  \quad \quad & \quad \quad & \quad \quad & \quad \quad \\
  \cline{1-4}
  \quad \quad & \quad \quad & \quad \quad & \quad \quad \\
  \cline{1-4}
  \quad \quad & \quad \quad & \quad \quad & \quad \quad \\
  \cline{1-4}
  \multicolumn{4}{c}{$\mathbf{1}(S=0)$}
\end{tabular}$ ,
$\begin{tabular}{|c|c|c|c|c|}
  \cline{1-5}
  \quad \quad & \quad \quad & \quad \quad & \quad \quad & \quad \quad  \\
  \cline{1-5}
  \quad \quad & \quad \quad & \quad \quad & \quad \quad \\
  \cline{1-4}
  \quad \quad & \quad \quad & \quad \quad \\
  \cline{1-3}
  \multicolumn{5}{c}{$\mathbf{8}(S=1)$}
\end{tabular}$,
$\begin{tabular}{|c|c|c|c|c|}
  \hline
  \quad \quad & \quad \quad & \quad \quad & \quad \quad & \quad \quad  \\
  \hline
  \quad \quad & \quad \quad & \quad \quad & \quad \quad & \quad \quad  \\
  \hline
  \quad \quad & \quad \quad \\
  \cline{1-2}
  \multicolumn{5}{c}{$\mathbf{\overline{10}}(S=1)$}
\end{tabular}$,
$\begin{tabular}{|c|c|c|c|c|c|}
  \hline
  \quad \quad & \quad \quad & \quad \quad & \quad \quad & \quad \quad & \quad \quad  \\
  \hline
  \quad \quad & \quad \quad & \quad \quad \\
  \cline{1-3}
  \quad \quad & \quad \quad & \quad \quad \\
  \cline{1-3}
  \multicolumn{6}{c}{$\mathbf{10}(S=1)$}
\end{tabular}$,
$\begin{tabular}{|c|c|c|c|c|c|}
  \hline
  \quad \quad & \quad \quad & \quad \quad & \quad \quad & \quad \quad & \quad \quad \\
  \hline
  \quad \quad & \quad \quad & \quad \quad & \quad \quad \\
  \cline{1-4}
  \quad \quad & \quad \quad \\
  \cline{1-2}
  \multicolumn{6}{c}{$\mathbf{27}(S=0)$}
\end{tabular}$.
\end{small}

\subsubsection{Three baryons + three diquarks}

The color and flavor $\otimes$ spin states of 15 quarks are as follows.\\
\\
Color : $\begin{tabular}{|c|c|c|c|c|}
  \hline
  \quad \quad & \quad \quad & \quad \quad & \quad \quad & \quad \quad \\
  \hline
  \quad \quad & \quad \quad & \quad \quad & \quad \quad & \quad \quad \\
  \hline
  \quad \quad & \quad \quad & \quad \quad & \quad \quad & \quad \quad \\
  \hline
\end{tabular}$, \quad
Flavor $\otimes$ Spin : $\begin{tabular}{|c|c|c|}
  \hline
  \quad \quad & \quad \quad & \quad \quad  \\
  \hline
  \quad \quad & \quad \quad & \quad \quad  \\
  \hline
  \quad \quad & \quad \quad & \quad \quad  \\
  \hline
  \quad \quad & \quad \quad & \quad \quad  \\
  \hline
  \quad \quad & \quad \quad & \quad \quad  \\
  \hline
\end{tabular}.$\\

Using the decomposition in octet baryon case, we can determine the possible flavor and spin states as follows.\\

Flavor states of 15 quarks :
\begin{align}
\mathbf{8} \times \mathbf{8} \times \mathbf{8} \times \mathbf{1} =& \mathbf{1}_{(m=2)} + \mathbf{8}_{(m=8)} + \mathbf{10}_{(m=4)} + \mathbf{\overline{10}}_{(m=4)} \nonumber \\
& + \mathbf{27}_{(m=6)} + \mathbf{35}_{(m=2)} + \mathbf{\overline{35}}_{(m=2)} + \mathbf{64}.
\end{align}

Now, we can determine the possible flavor and spin states as follows.\\

Flavor and spin :\\

$\begin{tabular}{|c|c|c|c|c|c|}
  \hline
  \quad \quad & \quad \quad & \quad \quad & \quad \quad & \quad \quad & \quad \quad \\
  \hline
  \quad \quad & \quad \quad & \quad \quad & \quad \quad & \quad \quad \\
  \cline{1-5}
  \quad \quad & \quad \quad & \quad \quad & \quad \quad \\
  \cline{1-4}
  \multicolumn{6}{c}{$\mathbf{8}(S=\frac{1}{2})$}
\end{tabular}$,
$\begin{tabular}{|c|c|c|c|c|c|}
  \hline
  \quad \quad & \quad \quad & \quad \quad & \quad \quad & \quad \quad & \quad \quad \\
  \hline
  \quad \quad & \quad \quad & \quad \quad & \quad \quad & \quad \quad & \quad \quad \\
  \hline
  \quad \quad & \quad \quad & \quad \quad \\
  \cline{1-3}
  \multicolumn{6}{c}{$\mathbf{\overline{10}}(S=\frac{3}{2})$}
\end{tabular}$.

\subsubsection{Four baryons + three diquarks}

The color and flavor $\otimes$ spin states of 18 quarks consist of four baryons and three correlated diquarks are as follows.\\

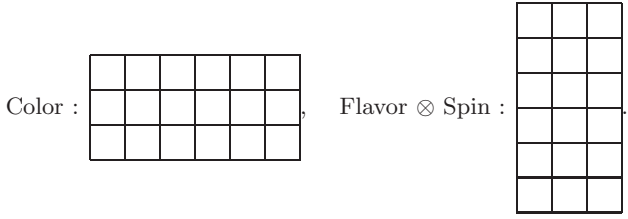
\begin{small}
Color : $\begin{tabular}{|c|c|c|c|c|c|}
  \hline
  \quad \quad & \quad \quad & \quad \quad & \quad \quad & \quad \quad & \quad \quad \\
  \hline
  \quad \quad & \quad \quad & \quad \quad & \quad \quad & \quad \quad & \quad \quad \\
  \hline
  \quad \quad & \quad \quad & \quad \quad & \quad \quad & \quad \quad & \quad \quad \\
  \hline
\end{tabular}$, \quad
Flavor $\otimes$ Spin : $\begin{tabular}{|c|c|c|}
  \hline
  \quad \quad & \quad \quad & \quad \quad  \\
  \hline
  \quad \quad & \quad \quad & \quad \quad  \\
  \hline
  \quad \quad & \quad \quad & \quad \quad  \\
  \hline
  \quad \quad & \quad \quad & \quad \quad  \\
  \hline
  \quad \quad & \quad \quad & \quad \quad  \\
  \hline
  \quad \quad & \quad \quad & \quad \quad  \\
  \hline
\end{tabular}.$
\end{small}\\

Flavor states of 18 quarks :
\begin{align}
  \mathbf{8} \times \mathbf{8} \times \mathbf{8} \times \mathbf{8} =& \mathbf{1}_{(m=8)}+\mathbf{8}_{(m=32)}+\mathbf{10}_{(m=20)}+\mathbf{\overline{10}}_{(m=20)} \nonumber \\
  & +\mathbf{27}_{(m=33)}+\mathbf{28}_{(m=2)}+\mathbf{\overline{28}}_{(m=2)} \nonumber \\
  &+\mathbf{35}_{(m=15)}+\mathbf{\overline{35}}_{(m=15)} +\mathbf{64}_{(m=12)} \nonumber\\
  & +\mathbf{81}_{(m=3)} +\overline{\mathbf{81}}_{(m=3)} +\mathbf{125}.
\end{align}

Using the decomposition in octet baryon case, we can find that there is one possible flavor and spin states as follows.\\

Flavor and spin :
$\begin{tabular}{|c|c|c|c|c|c|}
  \hline
  \quad \quad & \quad \quad & \quad \quad & \quad \quad & \quad \quad & \quad \quad \\
  \hline
  \quad \quad & \quad \quad & \quad \quad & \quad \quad & \quad \quad & \quad \quad \\
  \hline
  \quad \quad & \quad \quad & \quad \quad & \quad \quad & \quad \quad & \quad \quad \\
  \hline
  \multicolumn{6}{c}{$\mathbf{1}(S=0)$}
\end{tabular}$.

\subsubsection{Five baryons + three diquarks}

There is no allowed states satisfying the Pauli principle.

\subsection{Baryon(octet) case}

Now, we consider the case that a probe is a flavor octet baryon. Since we can refer the multiquark states composed of baryons only in Sec.\ref{multibaryon}, here we represent the five baryons plus one baryon case only.

\subsubsection{Five baryons + one baryon}
\label{5b1b}

The color and flavor $\otimes$ spin states of 18 quarks configuration are as follows.\\

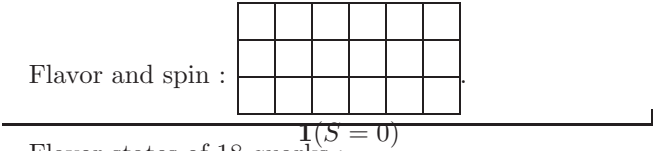
\begin{small}
Color : $\begin{tabular}{|c|c|c|c|c|c|}
  \hline
  \quad \quad & \quad \quad & \quad \quad & \quad \quad & \quad \quad & \quad \quad \\
  \hline
  \quad \quad & \quad \quad & \quad \quad & \quad \quad & \quad \quad & \quad \quad \\
  \hline
  \quad \quad & \quad \quad & \quad \quad & \quad \quad & \quad \quad & \quad \quad \\
  \hline
\end{tabular}$, \
Flavor $\otimes$ Spin : $\begin{tabular}{|c|c|c|}
  \hline
  \quad \quad & \quad \quad & \quad \quad  \\
  \hline
  \quad \quad & \quad \quad & \quad \quad  \\
  \hline
  \quad \quad & \quad \quad & \quad \quad  \\
  \hline
  \quad \quad & \quad \quad & \quad \quad  \\
  \hline
  \quad \quad & \quad \quad & \quad \quad  \\
  \hline
  \quad \quad & \quad \quad & \quad \quad \\
  \hline
\end{tabular}$.
\end{small}\\
The flavor $\otimes$ spin coupling state [3,3,3,3,3,3] with SU(6) can be decomposed into the states with the flavor SU(3) and the spin SU(2) as follows.
\begin{align}
[3,3,3,3,3,3]_{FS} &=[6,6,6]_F \otimes [9,9]_S
\end{align}

\begin{widetext}
Flavor states of 18 quarks :
\begin{align}
  \mathbf{8}\times \mathbf{8}\times \mathbf{8} \times \mathbf{8} \times \mathbf{8} \times \mathbf{8}
  =& \mathbf{1}_{(m=145)}+\mathbf{8}_{(m=702)}+\mathbf{10}_{(m=525)}+\overline{\mathbf{10}}_{(m=525)}+ \mathbf{27}_{(m=999)}+\mathbf{28}_{(m=161)}
+ \overline{\mathbf{28}}_{(m=161)} \nonumber\\
&+\mathbf{35}_{(m=630)} + \overline{\mathbf{35}}_{(m=630)}+ \mathbf{55}_{(m=5)} +\overline{\mathbf{55}}_{(m=5)} +\mathbf{64}_{(m=660)} +\mathbf{80}_{(m=70)} +\overline{\mathbf{80}}_{(m=70)}  \nonumber\\
& +\mathbf{81}_{(m=315)} +\overline{\mathbf{81}}_{(m=315)} +\mathbf{125}_{(m=215)} +\mathbf{154}_{(m=70)} +\overline{\mathbf{154}}_{(m=70)} +\mathbf{162}_{(m=9)} +\overline{\mathbf{162}}_{(m=9)} \nonumber\\
&+\mathbf{216}_{(m=30)} +\mathbf{260}_{(m=5)} +\overline{\mathbf{260}}_{(m=5)} +\mathbf{343}.
\end{align}
\end{widetext}

Flavor and spin : $\begin{tabular}{|c|c|c|c|c|c|}
  \hline
  \quad \quad & \quad \quad & \quad \quad & \quad \quad & \quad \quad & \quad \quad \\
  \hline
  \quad \quad & \quad \quad & \quad \quad & \quad \quad & \quad \quad & \quad \quad \\
  \hline
  \quad \quad & \quad \quad & \quad \quad & \quad \quad & \quad \quad & \quad \quad \\
  \hline
  \multicolumn{6}{c}{$\mathbf{1}(S=0)$}
\end{tabular}$.\\

\subsection{Baryon(decuplet) case}

In this subsection, we additionally consider the flavor decuplet baryon case. In Ref.\cite{MarquesL:2018fph}, we studied the delta isobar in normal nuclear matter using a constituent quark model. Investigating the stability of the decuplet baryon in dense matter could be interesting subject, so we include this case here.

\subsubsection{one baryon + one baryon(decuplet)}

Let's consider the 6 quarks state which is composed of one flavor octet baryon and one flavor decuplet baryon. The color and flavor $\otimes$ spin states are same as in Sec.\ref{2b}. However, the possible flavor and spin states are different after the decomposition, we need the following outer product.\\

Flavor states of 6 quarks :
\begin{align}
  \mathbf{8}\times \mathbf{10} = \mathbf{35} + \mathbf{27} + \mathbf{10} + \mathbf{8}.
\end{align}
Then, using the decomposition in octet baryon case, we can determine the possible flavor and spin states as follows.

Flavor and spin :
$\begin{tabular}{|c|c|c|c}
  \cline{1-3}
  \quad \quad & \quad \quad & \quad \quad \\
  \cline{1-3}
  \quad \quad & \quad \quad \\
  \cline{1-2}
  \quad \quad \\
  \cline{1-1}
  \multicolumn{4}{c}{$\mathbf{8}(S=1,2)$}
\end{tabular}$,
$\begin{tabular}{|c|c|c|c|c}
  \cline{1-4}
  \quad \quad & \quad \quad & \quad \quad & \quad \quad \\
  \cline{1-4}
  \quad \quad \\
  \cline{1-1}
  \quad \quad \\
  \cline{1-1}
  \multicolumn{5}{c}{$\mathbf{10}(S=1)$}
\end{tabular}$\quad ,
$\begin{tabular}{|c|c|c|c|c}
  \cline{1-4}
  \quad \quad & \quad \quad & \quad \quad & \quad \quad \\
  \cline{1-4}
  \quad \quad & \quad \quad \\
  \cline{1-2}
  \multicolumn{5}{c}{$\mathbf{27}(S=0)$}
\end{tabular}$\quad,
$\begin{tabular}{|c|c|c|c|c|}
  \hline
  \quad \quad & \quad \quad & \quad \quad & \quad \quad & \quad \quad \\
  \hline
  \quad \quad \\
  \cline{1-1}
  \multicolumn{5}{c}{$\mathbf{35}(S=1)$}
\end{tabular}$.

\subsubsection{two baryons + one baryon(decuplet)}

The color and flavor $\otimes$ spin states are same as in Sec.\ref{3b}. \\

Flavor states of 9 quarks :
\begin{align}
\mathbf{8}\times \mathbf{8}\times \mathbf{10} =& \mathbf{1} + \mathbf{8}_{(m=4)} + \mathbf{10}_{(m=4)} + \overline{\mathbf{10}}_{(m=2)} +\mathbf{27}_{(m=5)} \nonumber \\
& + \mathbf{28} + \mathbf{35}_{(m=4)} + \overline{\mathbf{35}} + \mathbf{64}_{(m=2)} + \mathbf{81}.
\end{align}

\begin{small}
Flavor and spin :
$\begin{tabular}{|c|c|c|c|c|}
  \cline{1-3}
  \quad \quad & \quad \quad & \quad \quad  \\
  \cline{1-3}
  \quad \quad & \quad \quad & \quad \quad \\
  \cline{1-3}
  \quad \quad & \quad \quad & \quad \quad \\
  \cline{1-3}
  \multicolumn{5}{c}{$\mathbf{1}(S=\frac{3}{2},\frac{5}{2})$}
\end{tabular}$,
$\begin{tabular}{|c|c|c|c|c|}
  \cline{1-4}
  \quad \quad & \quad \quad & \quad \quad & \quad \quad  \\
  \cline{1-4}
  \quad \quad & \quad \quad & \quad \quad \\
  \cline{1-3}
  \quad \quad & \quad \quad  \\
  \cline{1-2}
  \multicolumn{5}{c}{$\mathbf{8}(S=\frac{1}{2},\frac{3}{2},\frac{5}{2})$}
\end{tabular}$,
$\begin{tabular}{|c|c|c|c|c|}
  \hline
  \quad \quad & \quad \quad & \quad \quad & \quad \quad & \quad \quad \\
  \hline
  \quad \quad & \quad \quad  \\
  \cline{1-2}
  \quad \quad & \quad \quad \\
  \cline{1-2}
  \multicolumn{4}{c}{$\mathbf{10}(S=\frac{3}{2})$}
\end{tabular}$,
$\begin{tabular}{|c|c|c|c|}
  \hline
  \quad \quad & \quad \quad & \quad \quad & \quad \quad  \\
  \hline
  \quad \quad & \quad \quad & \quad \quad & \quad \quad  \\
  \hline
  \quad \quad \\
  \cline{1-1}
  \multicolumn{4}{c}{$\mathbf{\overline{10}}(S=\frac{3}{2})$}
\end{tabular}$,
$\begin{tabular}{|c|c|c|c|c|}
  \hline
  \quad \quad & \quad \quad & \quad \quad & \quad \quad & \quad \quad   \\
  \hline
  \quad \quad & \quad \quad & \quad \quad \\
  \cline{1-3}
  \quad \quad  \\
  \cline{1-1}
  \multicolumn{5}{c}{$\mathbf{27}(S=\frac{1}{2},\frac{3}{2},\frac{5}{2})$}
\end{tabular}$,
$\begin{tabular}{|c|c|c|c|c|c|}
  \hline
  \quad \quad & \quad \quad & \quad \quad & \quad \quad & \quad \quad & \quad \quad  \\
  \hline
  \quad \quad & \quad \quad   \\
  \cline{1-2}
  \quad \quad  \\
  \cline{1-1}
  \multicolumn{4}{c}{$\mathbf{35}(S=\frac{1}{2})$}
\end{tabular}$,
$\begin{tabular}{|c|c|c|c|c|}
  \cline{1-5}
  \quad \quad & \quad \quad & \quad \quad & \quad \quad & \quad \quad  \\
  \cline{1-5}
  \quad \quad & \quad \quad & \quad \quad & \quad \quad \\
  \cline{1-4}
  \multicolumn{4}{c}{$\overline{\mathbf{35}}(S=\frac{1}{2})$}
\end{tabular}$,
$\begin{tabular}{|c|c|c|c|c|c|}
  \cline{1-6}
  \quad \quad & \quad \quad & \quad \quad & \quad \quad & \quad \quad & \quad \quad \\
  \cline{1-6}
  \quad \quad & \quad \quad & \quad \quad \\
  \cline{1-3}
  \multicolumn{4}{c}{$\mathbf{64}(S=\frac{3}{2})$}
\end{tabular}$.
\end{small}

\subsubsection{three baryons + one baryon(decuplet)}

The color and flavor $\otimes$ spin states are same as in Sec.\ref{4b}. \\
\begin{widetext}
Flavor states of 12 quarks :
\begin{align}
\mathbf{8}\times \mathbf{8}\times \mathbf{8}\times \mathbf{10} =& \mathbf{1}_{(m=4)} + \mathbf{8}_{(m=20)} + \mathbf{10}_{(m=17)} + \overline{\mathbf{10}}_{(m=12)} +\mathbf{27}_{(m=27)} + \mathbf{28}_{(m=6)} + \overline{\mathbf{28}} + \mathbf{35}_{(m=21)} + \overline{\mathbf{35}}_{(m=11)}  \nonumber\\
 & + \mathbf{64}_{(m=15)} + \mathbf{80}_{(m=2)} + \mathbf{81}_{(m=9)} \overline{\mathbf{81}}_{(m=3)} + \mathbf{125}_{(m=3)} + \mathbf{154}.
\end{align}

Using the decomposition in octet baryon case, we can determine the possible flavor and spin states as follows.\\

Flavor and spin : $\begin{tabular}{|c|c|c|c|c|c|}
  \cline{1-6}
  \quad \quad & \quad \quad & \quad \quad & \quad \quad & \quad \quad & \quad \quad \\
  \cline{1-6}
  \quad \quad & \quad \quad & \quad \quad & \quad \quad & \quad \quad & \quad \quad \\
  \cline{1-6}
  \multicolumn{6}{c}{$\overline{\mathbf{28}}(S=0)$}
\end{tabular}$ ,
$\begin{tabular}{|c|c|c|c|c|c|}
  \cline{1-6}
  \quad \quad & \quad \quad & \quad \quad & \quad \quad & \quad \quad & \quad \quad  \\
  \cline{1-6}
  \quad \quad & \quad \quad & \quad \quad & \quad \quad & \quad \quad \\
  \cline{1-5}
  \quad \quad \\
  \cline{1-1}
  \multicolumn{6}{c}{$\overline{\mathbf{35}}(S=1)$}
\end{tabular}$,
$\begin{tabular}{|c|c|c|c|c|c|}
  \hline
  \quad \quad & \quad \quad & \quad \quad & \quad \quad & \quad \quad & \quad \quad  \\
  \hline
  \quad \quad & \quad \quad & \quad \quad & \quad \quad   \\
  \cline{1-4}
  \quad \quad & \quad \quad \\
  \cline{1-2}
  \multicolumn{6}{c}{$\mathbf{27}(S=0,2)$}
\end{tabular}$,
$\begin{tabular}{|c|c|c|c|c|c|}
  \hline
  \quad \quad & \quad \quad & \quad \quad & \quad \quad & \quad \quad & \quad \quad   \\
  \hline
  \quad \quad & \quad \quad & \quad \quad \\
  \cline{1-3}
  \quad \quad & \quad \quad & \quad \quad \\
  \cline{1-3}
  \multicolumn{6}{c}{$\mathbf{27}(S=1,3)$}
\end{tabular}$,
$\begin{tabular}{|c|c|c|c|c|}
  \hline
  \quad \quad & \quad \quad & \quad \quad & \quad \quad & \quad \quad \\
  \hline
  \quad \quad & \quad \quad & \quad \quad & \quad \quad & \quad \quad \\
  \cline{1-5}
  \quad \quad & \quad \quad \\
  \cline{1-2}
  \multicolumn{5}{c}{$\overline{\mathbf{10}}(S=1)$}
\end{tabular}$,
$\begin{tabular}{|c|c|c|c|c|}
  \hline
  \quad \quad & \quad \quad & \quad \quad & \quad \quad & \quad \quad  \\
  \hline
  \quad \quad & \quad \quad & \quad \quad & \quad \quad  \\
  \cline{1-4}
  \quad \quad & \quad \quad & \quad \quad \\
  \cline{1-3}
  \multicolumn{4}{c}{$\mathbf{8}(S=1,2)$}
\end{tabular}$,
$\begin{tabular}{|c|c|c|c|}
  \cline{1-4}
  \quad \quad & \quad \quad & \quad \quad & \quad \quad  \\
  \cline{1-4}
  \quad \quad & \quad \quad & \quad \quad & \quad \quad \\
  \cline{1-4}
  \quad \quad & \quad \quad & \quad \quad & \quad \quad \\
  \cline{1-4}
  \multicolumn{4}{c}{$\mathbf{1}(S=0)$}
\end{tabular}$.
\end{widetext}

\subsubsection{Four baryons + one baryon(decuplet)}
\label{4b1b2}

The color and flavor $\otimes$ spin states are same as in Sec.\ref{5b}. \\
\begin{widetext}
Flavor states of 15 quarks :
\begin{align}
  \mathbf{8} \times \mathbf{8} \times \mathbf{8} \times \mathbf{8} \times \mathbf{10}
  =& \mathbf{1}_{(m=20)}+\mathbf{8}_{(m=100)}+\mathbf{10}_{(m=85)}+\mathbf{\overline{10}}_{(m=70)}+\mathbf{27}_{(m=150)}+\mathbf{28}_{(m=38)}+\mathbf{\overline{28}}_{(m=15)} +\mathbf{35}_{(m=116)} \nonumber\\
   &+\mathbf{\overline{35}}_{(m=80)} +\mathbf{55}_{(m=2)} +\mathbf{64}_{(m=104)} +\mathbf{80}_{(m=20)} +\overline{\mathbf{80}}_{(m=4)} +\mathbf{81}_{(m=66)} +\overline{\mathbf{81}}_{(m=36)} +\mathbf{125}_{(m=34)} \nonumber\\
   & +\mathbf{154}_{(m=16)} +\overline{\mathbf{154}}_{(m=6)} +\mathbf{162}_{(m=3)} +\mathbf{216}_{(m=4)} + \mathbf{260}.
\end{align}

Flavor and spin :
$\begin{tabular}{|c|c|c|c|c|c|}
  \hline
  \quad \quad & \quad \quad & \quad \quad & \quad \quad & \quad \quad & \quad \quad \\
  \hline
  \quad \quad & \quad \quad & \quad \quad & \quad \quad & \quad \quad & \quad \quad \\
  \hline
  \quad \quad & \quad \quad & \quad \quad \\
  \cline{1-3}
  \multicolumn{5}{c}{$\mathbf{\bar{10}}(S=\frac{3}{2})$}
\end{tabular}$,
$\begin{tabular}{|c|c|c|c|c|c|}
  \hline
  \quad \quad & \quad \quad & \quad \quad & \quad \quad & \quad \quad & \quad \quad \\
  \hline
  \quad \quad & \quad \quad & \quad \quad & \quad \quad & \quad \quad \\
  \cline{1-5}
  \quad \quad & \quad \quad & \quad \quad & \quad \quad \\
  \cline{1-4}
  \multicolumn{6}{c}{$\mathbf{8}(S=\frac{1}{2})$}
\end{tabular}$.
\end{widetext}

\subsubsection{Five baryons + one baryon(decuplet)}
\label{5b1b2}

The color and flavor $\otimes$ spin states are same as in Sec.\ref{5b1b}. \\
\begin{widetext}
Flavor states of 18 quarks :
\begin{align}
  \mathbf{8}\times \mathbf{8}\times \mathbf{8} \times \mathbf{8} \times \mathbf{8} \times \mathbf{10}
  =& \mathbf{1}_{(m=100)}+\mathbf{8}_{(m=525)}+\mathbf{10}_{(m=451)}+\overline{\mathbf{10}}_{(m=400)}+ \mathbf{27}_{(m=855)}+\mathbf{28}_{(m=240)}
+ \overline{\mathbf{28}}_{(m=135)} \nonumber\\
& +\mathbf{35}_{(m=675)} + \overline{\mathbf{35}}_{(m=535)}+ \mathbf{55}_{(m=25)} +\overline{\mathbf{55}}_{(m=4)} +\mathbf{64}_{(m=690)} +\mathbf{80}_{(m=165)} +\overline{\mathbf{80}}_{(m=65)}  \nonumber\\
& +\mathbf{81}_{(m=460)} +\overline{\mathbf{81}}_{(m=315)} +\mathbf{125}_{(m=300)} +\mathbf{143}_{(m=5)} +\mathbf{154}_{(m=160)} +\overline{\mathbf{154}}_{(m=90)}  \nonumber\\
& +\mathbf{162}_{(m=45)} +\overline{\mathbf{162}}_{(m=10)}+\mathbf{216}_{(m=65)} +\mathbf{260}_{(m=25)} +\overline{\mathbf{260}}_{(m=10)} +\mathbf{280}_{(m=4)} +\mathbf{343}_{(m=5)} \nonumber\\
& +\mathbf{405}.
\end{align}
\end{widetext}
Flavor and spin : $\begin{tabular}{|c|c|c|c|c|c|}
  \hline
  \quad \quad & \quad \quad & \quad \quad & \quad \quad & \quad \quad & \quad \quad \\
  \hline
  \quad \quad & \quad \quad & \quad \quad & \quad \quad & \quad \quad & \quad \quad \\
  \hline
  \quad \quad & \quad \quad & \quad \quad & \quad \quad & \quad \quad & \quad \quad \\
  \hline
  \multicolumn{6}{c}{$\mathbf{1}(S=0)$}
\end{tabular}$.\\

\section{Free quark gas}
\label{free-quark}

Finally, we consider the case where the surrounding is a free quark gas. In such a case, we assume that the surrounding free quarks are not correlated with each other, but are correlated with the probe to satisfy the Pauli principle.

\subsection{Quark case}

When a probe is a quark, we only need to calculate the average value of the color-spin interactions for all possible diquark configurations.
There are four diquark states satisfying the Pauli principle. We represent it in the Table \ref{two-quark-interaction}. If we compare it with the results for baryons and a quark, then we should multiply it by 3 to ensure comparison at the same density.

\begin{table}
\begin{center}
\begin{tabular}{c|c|c|c|c}
\hline
\hline
& \multicolumn{4}{c}{$q_i q_j$} \\
\hline
Flavor & $A$ & $S$ & $A$ &$S$ \\
\hline
Color & $A(\overline{3})$ & $A(\overline{3})$  & $S(6)$  & $S(6)$ \\
\hline
Spin & $A(1)$ & $S(3)$ & $S(3)$ & $A(1)$ \\
\hline
$- \lambda_i^c \lambda_j^c \sigma_i \cdot \sigma_j$ & $-8$& $\frac{8}{3}$ &  $-\frac{4}{3}$ & $4$ \\
\hline
$ \lambda_i^c \lambda_j^c $ & $-\frac{8}{3}$& $-\frac{8}{3}$ &  $\frac{4}{3}$ & $\frac{4}{3}$ \\
\hline
\end{tabular}
\end{center}
\caption{Classification of two quark interaction due to the Pauli exclusion principle. We denote the antisymmetric and symmetric state as $A$ and $S$, respectively. The symbols inside the parenthesis represent the  multiplet state.}
\label{two-quark-interaction}
\end{table}

\subsection{Diquark$(C_A,F_A,S_A)$ case}
\label{dq1}

For a diquark with color antitriplet and a free quark, since $\mathbf{3}\times \overline{\mathbf{3}}= \mathbf{1} + \mathbf{8}$, there are two possible color states of three quarks, which will come with the flavor $\otimes$ spin configuration as below.

1. Color : $\begin{tabular}{|c|}
  \hline
  \quad \quad \\
  \hline
  \quad \quad \\
  \hline
  \quad \quad \\
  \hline
\end{tabular}$, \quad
Flavor $\otimes$ Spin : $\begin{tabular}{|c|c|c|}
  \hline
  \quad \quad & \quad \quad & \quad \quad  \\
  \hline
\end{tabular}.$\\

The flavor $\otimes$ spin coupling state [3] with SU(6) can be decomposed into the states with the flavor SU(3) and the spin SU(2) as follows.
\begin{align}
[3]_{FS} =[3]_F \otimes [3]_S + [2,1]_F \otimes [2,1]_S.
\end{align}

Flavor and spin :
$\begin{tabular}{|c|c|c}
  \cline{1-2}
  \quad \quad & \quad \quad \\
  \cline{1-2}
  \quad \quad \\
  \cline{1-1}
  \multicolumn{3}{c}{$\mathbf{8}(S=\frac{1}{2})$}
\end{tabular}$.\\

2. Color : $\begin{tabular}{|c|c|}
  \hline
  \quad \quad & \quad \quad \\
  \hline
  \quad \quad \\
  \cline{1-1}
\end{tabular}$, \quad
Flavor $\otimes$ Spin : $\begin{tabular}{|c|c|}
  \hline
  \quad \quad & \quad \quad \\
  \hline
  \quad \quad \\
  \cline{1-1}
\end{tabular}.$\\

The flavor $\otimes$ spin coupling state [2,1] with SU(6) can be decomposed into the states with the flavor SU(3) and the spin SU(2) as follows.
\begin{align}
[2,1]_{FS} =& [3]_F \otimes [2,1]_S + [2,1]_F \otimes [3]_S + [2,1]_F \otimes [2,1]_S \nonumber \\
&+ [1,1,1]_F \otimes [2,1]_S.
\end{align}

Flavor and spin :
$\begin{tabular}{|c|c}
  \cline{1-1}
  \quad \quad \\
  \cline{1-1}
  \quad \quad \\
  \cline{1-1}
  \quad \quad \\
  \cline{1-1}
  \multicolumn{2}{c}{$\mathbf{1}(S=\frac{1}{2})$}
\end{tabular}$,
$\begin{tabular}{|c|c|c}
  \cline{1-2}
  \quad \quad & \quad \quad \\
  \cline{1-2}
  \quad \quad \\
  \cline{1-1}
  \multicolumn{3}{c}{$\mathbf{8}(S=\frac{1}{2})$}
\end{tabular}$.\\

Considering both cases, we can calculate the average value of the color-spin interaction. We multiply it by $\frac{3}{2}$ to compare this results with that for baryons and a quark.

\subsection{Diquark$(C_A,F_S,S_S)$ case}

In this case, there are two possible color states as in Sec.\ref{dq1}. 

1. Color : $\begin{tabular}{|c|}
  \hline
  \quad \quad \\
  \hline
  \quad \quad \\
  \hline
  \quad \quad \\
  \hline
\end{tabular}$, \quad
Flavor $\otimes$ Spin : $\begin{tabular}{|c|c|c|}
  \hline
  \quad \quad & \quad \quad & \quad \quad  \\
  \hline
\end{tabular}.$\\

Flavor and spin :
$\begin{tabular}{|c|c|c}
  \cline{1-2}
  \quad \quad & \quad \quad \\
  \cline{1-2}
  \quad \quad \\
  \cline{1-1}
  \multicolumn{3}{c}{$\mathbf{8}(S=\frac{1}{2})$}
\end{tabular}$,
$\begin{tabular}{|c|c|c|c}
  \cline{1-3}
  \quad \quad & \quad \quad & \quad \quad \\
  \cline{1-3}
  \multicolumn{4}{c}{$\mathbf{10}(S=\frac{3}{2})$}
\end{tabular}$.\\

2. Color : $\begin{tabular}{|c|c|}
  \hline
  \quad \quad & \quad \quad \\
  \hline
  \quad \quad \\
  \cline{1-1}
\end{tabular}$, \quad
Flavor $\otimes$ Spin : $\begin{tabular}{|c|c|}
  \hline
  \quad \quad & \quad \quad \\
  \hline
  \quad \quad \\
  \cline{1-1}
\end{tabular}.$\\

Flavor and spin :
$\begin{tabular}{|c|c|c}
  \cline{1-2}
  \quad \quad & \quad \quad \\
  \cline{1-2}
  \quad \quad \\
  \cline{1-1}
  \multicolumn{3}{c}{$\mathbf{8}(S=\frac{1}{2},\frac{3}{2})$}
\end{tabular}$,
$\begin{tabular}{|c|c|c|c}
  \cline{1-3}
  \quad \quad & \quad \quad & \quad \quad \\
  \cline{1-3}
  \multicolumn{4}{c}{$\mathbf{10}(S=\frac{1}{2})$}
\end{tabular}$.\\

\subsection{Diquark$(C_S,F_A,S_S)$ case}

For a diquark with color sextet, since $\mathbf{6}\times \mathbf{3}= \mathbf{8} + \mathbf{10}$, there are two possible color states of three quarks, which will come with the flavor $\otimes$ spin configuration as below.

1. Color : $\begin{tabular}{|c|c|}
  \hline
  \quad \quad & \quad \quad \\
  \hline
  \quad \quad \\
  \cline{1-1}
\end{tabular}$, \quad
Flavor $\otimes$ Spin : $\begin{tabular}{|c|c|}
  \hline
  \quad \quad & \quad \quad \\
  \hline
  \quad \quad \\
  \cline{1-1}
\end{tabular}$.\\

Flavor and spin :
$\begin{tabular}{|c|c}
  \cline{1-1}
  \quad \quad \\
  \cline{1-1}
  \quad \quad \\
  \cline{1-1}
  \quad \quad \\
  \cline{1-1}
  \multicolumn{2}{c}{$\mathbf{1}(S=\frac{1}{2})$}
\end{tabular}$,
$\begin{tabular}{|c|c|c}
  \cline{1-2}
  \quad \quad & \quad \quad \\
  \cline{1-2}
  \quad \quad \\
  \cline{1-1}
  \multicolumn{3}{c}{$\mathbf{8}(S=\frac{1}{2},\frac{3}{2})$}
\end{tabular}$.\\

2. Color : $\begin{tabular}{|c|c|c|}
  \hline
  \quad \quad & \quad \quad & \quad \quad \\
  \hline
\end{tabular}$, \quad
Flavor $\otimes$ Spin : $\begin{tabular}{|c|}
  \hline
  \quad \quad \\
  \hline
  \quad \quad \\
  \hline
  \quad \quad \\
  \hline
\end{tabular}$.\\

The flavor $\otimes$ spin coupling state [1,1,1] with SU(6) can be decomposed into the states with the flavor SU(3) and the spin SU(2) as follows.
\begin{align}
[1,1,1]_{FS} =[2,1]_F \otimes [2,1]_S + [1,1,1]_F \otimes [3]_S.
\end{align}

Flavor and spin :
$\begin{tabular}{|c|c}
  \cline{1-1}
  \quad \quad \\
  \cline{1-1}
  \quad \quad \\
  \cline{1-1}
  \quad \quad \\
  \cline{1-1}
  \multicolumn{2}{c}{$\mathbf{1}(S=\frac{3}{2})$}
\end{tabular}$,
$\begin{tabular}{|c|c|c}
  \cline{1-2}
  \quad \quad & \quad \quad \\
  \cline{1-2}
  \quad \quad \\
  \cline{1-1}
  \multicolumn{3}{c}{$\mathbf{8}(S=\frac{1}{2})$}
\end{tabular}$.\\

\subsection{Diquark$(C_S,F_S,S_A)$ case}

Here as elsewhere, there are two possible color states. \\

1. Color : $\begin{tabular}{|c|c|}
  \hline
  \quad \quad & \quad \quad \\
  \hline
  \quad \quad \\
  \cline{1-1}
\end{tabular}$, \quad
Flavor $\otimes$ Spin : $\begin{tabular}{|c|c|}
  \hline
  \quad \quad & \quad \quad \\
  \hline
  \quad \quad \\
  \cline{1-1}
\end{tabular}$.\\

Flavor and spin :
$\begin{tabular}{|c|c|c}
  \cline{1-2}
  \quad \quad & \quad \quad \\
  \cline{1-2}
  \quad \quad \\
  \cline{1-1}
  \multicolumn{3}{c}{$\mathbf{8}(S=\frac{1}{2})$}
\end{tabular}$,
$\begin{tabular}{|c|c|c|c}
  \cline{1-3}
  \quad \quad & \quad \quad & \quad \quad \\
  \cline{1-3}
  \multicolumn{4}{c}{$\mathbf{10}(S=\frac{1}{2})$}
\end{tabular}$.\\

2. Color : $\begin{tabular}{|c|c|c|}
  \hline
  \quad \quad & \quad \quad & \quad \quad \\
  \hline
\end{tabular}$, \quad
Flavor $\otimes$ Spin : $\begin{tabular}{|c|}
  \hline
  \quad \quad \\
  \hline
  \quad \quad \\
  \hline
  \quad \quad \\
  \hline
\end{tabular}$.\\

Flavor and spin :
$\begin{tabular}{|c|c|c}
  \cline{1-2}
  \quad \quad & \quad \quad \\
  \cline{1-2}
  \quad \quad \\
  \cline{1-1}
  \multicolumn{3}{c}{$\mathbf{8}(S=\frac{1}{2})$}
\end{tabular}$.\\

\subsection{Three correlated diquarks case}

For the color-spin interaction between  three correlated diquarks and a free quark, the color and flavor $\otimes$ spin coupling state is as follows.\\
\\
Color : $\begin{tabular}{|c|c|c|}
  \hline
  \quad \quad & \quad \quad & \quad \quad \\
  \hline
  \quad \quad & \quad \quad \\
  \cline{1-2}
  \quad \quad & \quad \quad \\
  \cline{1-2}
\end{tabular}$, \quad
Flavor $\otimes$ Spin : $\begin{tabular}{|c|c|c|}
  \hline
  \quad \quad & \quad \quad & \quad \quad  \\
  \hline
  \quad \quad & \quad \quad & \quad \quad  \\
  \hline
  \quad \quad \\
  \cline{1-1}
\end{tabular}$.\\

Flavor states of 7 quarks : $
\mathbf{1}\times \mathbf{3} = \mathbf{3}$. \\

Flavor and spin :
$\begin{tabular}{|c|c|c|}
  \hline
  \quad \quad & \quad \quad & \quad \quad \\
  \hline
  \quad \quad & \quad \quad \\
  \cline{1-2}
  \quad \quad & \quad \quad \\
  \cline{1-2}
  \multicolumn{3}{c}{$\mathbf{3}(S=\frac{1}{2})$}
\end{tabular}$.\\

In order to compare the result with the interaction factor when one baryon looks at one quark, we need to divide by 2.

\subsection{Baryon(octet) case}

The multiquark state of this case is the same as in Sec.\ref{1b1q}.

\subsection{Baryon(decuplet) case}

The color and flavor $\otimes$ spin states of 4 quarks configuration are as follows.\\

Color : $\begin{tabular}{|c|c|}
  \cline{1-2}
  \quad \quad & \quad \quad   \\
  \cline{1-2}
  \quad \quad \\
  \cline{1-1}
  \quad \quad \\
  \cline{1-1}
\end{tabular}$, \quad
Flavor $\otimes$ Spin : $\begin{tabular}{|c|c|c|}
  \cline{1-3}
  \quad \quad & \quad \quad & \quad \quad  \\
  \cline{1-3}
  \quad \quad \\
  \cline{1-1}
\end{tabular}$.\\

Flavor states of 4 quarks : $
\mathbf{10}\times \mathbf{3} = \mathbf{15'} + \mathbf{15}$. \\

Flavor and spin :
$\begin{tabular}{|c|c|c|c|}
  \hline
  \quad \quad & \quad \quad & \quad \quad & \quad \quad \\
  \hline
  \multicolumn{4}{c}{$\mathbf{15'}(S=1)$}
\end{tabular}$,
$\begin{tabular}{|c|c|c|c}
  \cline{1-3}
  \quad \quad & \quad \quad & \quad \quad \\
  \cline{1-3}
  \quad \quad \\
  \cline{1-1}
  \multicolumn{4}{c}{$\mathbf{15}(S=1,2)$}
\end{tabular}$.\\

\section{Results}
\label{results}

\begin{widetext}
\begin{table*}
\begin{tabular}{|c|c|c|c|c|c|c|}
  \hline
  SU(2)$_F$ &  1b & 2b & 3b & 4b & 5b & Free quarks \\
  \hline
  quark & 8 & 8.533 & 6.133 & None & None & 4.364 \\
  \hline
  diquark($C_A,F_A,S_A$) & 8 & 8 & 8 & None & None & 8 \\
  \hline
  diquark($C_A,F_S,S_S$) & 8 & 8.267 & 4.889 & None & None & 2.667 \\
  \hline
  diquark($C_S,F_A,S_S$) & 8 & 8.533 & None & None & None & 4.471 \\
  \hline
  diquark($C_S,F_S,S_A$) & 8 & 8.4 & None & None & None & 2.118 \\
  \hline
  baryon(octet) & 7.111 & 7.111 & 7.111 & None & None & 8 \\
  \hline
  baryon(decuplet) & 7.111 & 7.111 & 3.556 & None & None & 2.872 \\
  \hline
\end{tabular}
\begin{tabular}{|c|c|c|c|c|c|c|}
  \hline
  SU(3)$_F$ &  1b & 2b & 3b & 4b & 5b & Free quarks \\
  \hline
  quark & 6 & 6.446 & 4.644 & 4.167 & 3.657 & 2.823 \\
  \hline
  diquark($C_A,F_A,S_A$) & 6 & 6.176 & 5.551 & 5.257 & 4.8 & 6.3 \\
  \hline
  diquark($C_A,F_S,S_S$) & 6 & 6.107 & 3.799 & 3.359 & 2.933 & 1.12 \\
  \hline
  diquark($C_S,F_A,S_S$) & 6 & 6.4 & 4.884 & 4.376 & None & 3.28 \\
  \hline
  diquark($C_S,F_S,S_A$) & 6 & 6.185 & 3.869 & 3.304 & None & 0.643 \\
  \hline
  three correlated diquarks & 6 & 6 & 6 & 6 & None & 6 \\
  \hline
  baryon(octet) & 5.714 & 5.78 & 4.944 & 4.667 & 4.267 & 6 \\
  \hline
  baryon(decuplet) & 4.647 & 4.456 & 2.916 & 2.26 & 2.133 & 1.455 \\
  \hline
\end{tabular}\\
\caption{$\Delta H_{CS}^{\mathrm{avg}}$ for different probes (column) in  various surroundings (row) The upper and lower tables are for flavor SU(2) and SU(3), respectively. 'None' represents that there is no state that satisfies the Pauli principle.}
\label{interaction}
\end{table*}
\end{widetext}

Table \ref{interaction} shows the averaged color-spin interaction factors  calculated using the multiquark states and Eq.(\ref{2b-2}). In the case of flavor SU(2), the result was obtained by excluding the states that must contain a strange quark from among the possible states in flavor SU(3). In the case of flavor SU(2), since there is no strange quark, the combinations that make up an antisymmetric flavor state are quite limited. Therefore, from 4 surrounding baryons, it is not possible to create a state that satisfies the Pauli principle regardless of the probe type.

There are a few points on these results. First of all, as we can see in the Table \ref{interaction}, when the number of surrounding baryons is increased to two, the strength of the interaction increases, but from three it steadily decreases for most cases. Also, comparing the interaction between a quark and a octet baryon, it can be seen that quark is more repulsive when the number of surrounding baryons is up to two, but a baryon becomes more repulsive relatively from three onwards. It shows that the interaction experienced by a quark is relatively more attractive than that of a baryon once the  density increases to a point where three or more baryons are  correlated.  

It should be noted that ,in the case of three correlated diquarks, the averaged interaction factor does not differ when the number of surrounding baryons changes. The reason is that the states of the three correlated diquarks are all singlet, so when we multiply it by the multibaryon state, the multiplet state of multibaryon remains intact.

The results when the surrounding is a  free quark gas are also noteworthy. In the case of free quark gas, it can be seen that the interaction factor of quark is much more attractive than that of baryon as we expected. Additionally, for free quark gas, the interaction factor is more attractive compared to the case where the surroundings are multibaryon states. However, there are exceptions which will be explained below when discussing  the internal structure of a probe.

When we compare the octet baryon and the decuplet baryon, the interaction of the decuplet baryon seems more attractive. Even in the case of diquarks, it can be seen that the diquark with ($C_A,F_S,S_S$) is more attractive than the diquark with ($C_A,F_A,S_A$), which is considered the most attractive. However, these results are because of the internal color-spin factor of the probe. The diquark with ($C_A,F_S,S_S$) and the decuplet baryon have 
one thing in common: the flavor and spin states are totally symmetric. 

As we can find out in Eq.(\ref{CSF-1}), the color-spin factor is more attractive when there are more antisymmetric combination in the flavor state because the corresponding $C_F$ is small. The possibility that the surrounding baryons can form an antisymmetric combination with the probe increases when the flavor state of the probe is symmetric.
Then, the color-spin interaction between a probe and the surrounding baryons becomes attractive. However, in this case, we need to consider their internal structures because when the flavor state is symmetric then the color-spin factors are repulsive. Therefore, we show the results of considering the probe's internal color-spin factor as well as the interaction between the probe and the surrounding baryons in Table.\ref{energy}.

\begin{widetext}
\begin{table*}
\begin{tabular}{|c|c|c|c|c|c|c|}
  \hline
  SU(2)$_F$ &  1b & 2b & 3b & 4b & 5b & Free quarks \\
  \hline
  quark & 8 & 8.533 & 6.133 & None & None & 4.364 \\
  \hline
  diquark($C_A,F_A,S_A$) & 4 & 4 & 4 & None & None & 4 \\
  \hline
  diquark($C_A,F_S,S_S$) & 9.333 & 9.6 & 6.222 & None & None & 4 \\
  \hline
  diquark($C_S,F_A,S_S$) & 7.333 & 7.867 & None & None & None & 3.804 \\
  \hline
  diquark($C_S,F_S,S_A$) & 10 & 10.4 & None & None & None & 4.118 \\
  \hline
  baryon(octet) & 4.444 & 4.444 & 4.444 & None & None & 5.333 \\
  \hline
  baryon(decuplet) & 9.778 & 9.778 & 6.222 & None & None & 5.538 \\
  \hline
\end{tabular}\\
\begin{tabular}{|c|c|c|c|c|c|c|}
  \hline
  SU(3)$_F$ &  1b & 2b & 3b & 4b & 5b & Free quarks \\
  \hline
  quark & 6 & 6.446 & 4.644 & 4.167 & 3.657 & 2.823 \\
  \hline
  diquark($C_A,F_A,S_A$) & 2 & 2.176 & 1.551 & 1.257 & 0.8 & 2.4 \\
  \hline
  diquark($C_A,F_S,S_S$) & 7.333 & 7.44 & 5.132 & 4.692 & 4.267 & 2.45 \\
  \hline
  diquark($C_S,F_A,S_S$) & 5.333 & 5.734 & 4.217 & 3.709 & None & 2.613 \\
  \hline
  diquark($C_S,F_S,S_A$) & 8 & 8.185 & 5.869 & 5.304 & None & 2.643 \\
  \hline
  three correlated diquarks & 2 & 2 & 2 & 2 & None & 2 \\
  \hline
  baryon(octet) & 3.048 & 3.113 & 2.277 & 2 & 1.6 & 3.333 \\
  \hline
  baryon(decuplet) & 7.313 & 7.123 & 5.583 & 4.927 & 4.8 & 4.121 \\
  \hline
\end{tabular}
\caption{Same as the Table \ref{interaction} after the internal interactions within the probe are considered.}
\label{energy}
\end{table*}
\end{widetext}

Considering the internal color-spin factors of probes, 
it can be seen that when the surrounding is baryons, and the more baryons are correlated, the most stable state is the diquark with ($C_A,F_A,S_A$). Therefore, 
if a baryon is decomposed into a quark and a diquark and the accumulated quarks further form  diquarks 
it can be seen that this result is consistent with the so-called diquarkyonic matter configuration in which quark and diquark can coexist at high density.
Even at higher density when the surrounding turns into  the free quark gas, the most stable state is three correlated diquarks.

\section{summary}
\label{summary}

In this work, we constructed the multiquark states to calculate the interaction energy of a probe. To examine the behavior of dense matter, we calculated the relative magnitude of the interaction experienced by the probe as the number of correlated surrounding baryons increased but density kept constant. As a result, we found that a quark experiences the less repulsive interaction than a baryon when the number of correlated surrounding baryons is 3 or more. 

On the other hand, we showed that the diquark may be the most stable state in dense matter when the internal interactions of a probe are considered. These results show the possibility of a new phase called diquarkyonic matter. 

There are a few additional things that need to be done in relation to this. If diquark can exist as an independent state, then the interaction between diquarks cannot be ignored. 
Investigation on the interactions between diquarks can be an important factor in the study of diquark condensation as well as diquarkyonic configuration. Additionally, it is also necessary to consider the inhomogeneous matter. It can be important to look at how the behavior of an interaction changes when the spatial part of a multiquark state is not totally symmetric.

\section*{Acknowledgments}
This work was supported by Samsung Science and Technology Foundation under Project Number SSTF-BA1901-04. The work of A.P. was supported by the Korea National Research Foundation under the grant number 2021R1I1A1A01043019. 

\begin{appendix}

\section{Three-body confinement potential in SU($N$)}

The commutation and anticommutation relations for generators of SU($N$) are as follow. 
\begin{align}
    [T^a,T^b] &= if^{abc}T^c, \\
    \{T^a, T^b \} &= \frac{1}{N}\delta^{ab} + d^{abc}T^c,
\end{align}
where $a,b,c=1,2,\cdots, N^2-1$. Normalization condition is as follows.
\begin{align}
    \text{Tr}(T^a T^b) = \frac{1}{2}\delta^{ab}.
\end{align}

We can also represent the color-color interaction between $i$'th and $j$'th quark using the permutation.
\begin{align}
    T_i^a T_j^a = \frac{1}{2}(ij) -\frac{1}{2N}I,
\end{align}
where $I$ is the $(N^2-1)$ by $(N^2-1)$ identity matrix. Then, we can represent the $d$-type and $f$-type three body confinement forces as follows.

\begin{align}
    d^{abc}T^a_i T^b_j T^c_k &= \frac{1}{4}[(ijk)+(ikj)] + 
    \frac{1}{N^2}I \nonumber \\
    & \quad -
    \frac{1}{2N}[(ij)+(ik)+(jk)],\\
    f^{abc}T^a_i T^b_j T^c_k &= \frac{i}{4}[(ijk)-(ikj)].
\end{align}
Now, consider the following general color state of SU($N$).\\

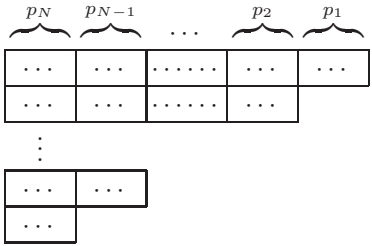
\begin{tabular}{|c|c|c|c|c|}
  \multicolumn{1}{c}{$\overbrace{\rule{.8cm}{0pt}}^{p_N}$} & \multicolumn{1}{c}{$\overbrace{\rule{.8cm}{0pt}}^{p_{N-1}}$} & \multicolumn{1}{c}{$\cdots$} & \multicolumn{1}{c}{$\overbrace{\rule{.8cm}{0pt}}^{p_2}$} & \multicolumn{1}{c}{$\overbrace{\rule{.8cm}{0pt}}^{p_1}$} \\
  \cline{1-5}
  $\cdots$ & $\cdots$ & $\cdots \cdots$ & $\cdots$ & $\cdots$ \\
  \cline{1-5}
  $\cdots$ & $\cdots$ & $\cdots \cdots$ & $\cdots$ \\
  \cline{1-4}
  \multicolumn{1}{c}{$\vdots$} \\
  \cline{1-2}
  $\cdots$ & $\cdots$ \\
  \cline{1-2}
  $\cdots$ \\
  \cline{1-1}
\end{tabular} 
\\

Since the $d$-type three-body confinement potential is Casimir operator, we can get the eigenvalue of it by calculating the diagonal component for the normal Young-Yamanouchi basis.

There are four cases. 1) $i,j,k$ are in the same row. 2) $i,j$ are in the same row and $k$ is in the lower row. 3) $j,k$ are in the same row and $i$ is in the upper row. 4) $i,j,k$ are in different rows. Here we show for  cases 1) and 2). For cases 3) and 4), we can calculate it in a similar way.

For case 1), since $i,j,k$ are in the same row, any permutation between $i,j,k$ is just the identity. And the number of this case is as follows.

\begin{align}
    N_1 &= \binom{p_1+ \cdots +p_N}{3} + \binom{p_2 + \cdots + p_N}{3} + \cdots + \binom{p_N}{3}.
\end{align}

For case 2), we can calculate it step by step. First, consider the case when $i,j$ are in the first row. Then the possible number of this case is as follows.

\begin{align}
    N_{2,1} &= \binom{p_1 + \cdots + p_N}{2} \left[ \binom{p_2 + \cdots +p_N}{1} + \cdots + \binom{p_N}{1} \right].
\end{align}

Also, the diagonal components of each permutation are as follows.

\begin{align}
    &(ij)=1, \quad (ik)=(jk)=-\frac{1}{p_1 + \cdots + p_N}, \nonumber \\
    &(ijk)=(ij)(jk)=(jk) =(ikj).
\end{align}

We can continue this calculation when $i,j$ are in the second row, third row, $\cdots$ and finally $(N-1)$'th row. The possible number of the final case and the diagonal components of each permutation are as follows.

\begin{align}
    N_{2,N-1} &= \binom{p_{N-1}+p_N}{2} \binom{p_N}{1}.
\end{align}

\begin{align}
    &(ij)=1, \quad (ik)=(jk)=-\frac{1}{p_{N-1}+p_N}, \nonumber\\
    &(ijk)=(ij)(jk)=(jk)=(ikj).
\end{align}

For cases 3) and 4) we can use similar method. Collecting all terms, we can get the eigenvalue of $d$-type three-body confinement force. Here, we represent it for SU(4), SU(5) and SU(6).

\begin{widetext}
For SU(4),
\begin{align}
    \sum_{i<j<k}d^{abc}T_i^a T_j^b T_k^c &= \frac{p_1}{16} -\frac{3p_1^2}{32} + \frac{p_1^3}{32} + \frac{3p_2}{8} -\frac{p_1 p_2}{4} + \frac{p_1^2 p_2}{16} -\frac{3p_2^2}{8} + \frac{23p_3}{16} - \frac{3p_1 p_3}{16} + \frac{p_1^2 p_3}{32} -\frac{p_2 p_3}{2} -\frac{15 p_3^2}{32} -\frac{p_1 p_3^2}{32} \nonumber \\
    &\quad   -\frac{p_2 p_3^2}{16} -\frac{p_3^3}{32} + \frac{15 p_4}{4}  
\end{align}

For SU(5), 
\begin{align}
    \sum_{i<j<k}d^{abc}T_i^a T_j^b T_k^c &= \frac{2p_1}{25} - \frac{3p_1^2}{25} + \frac{p_1^3}{25} + \frac{23p_2}{50} -\frac{33p_1 p_2}{100} + \frac{9p_1^2 p_2}{100} -\frac{12p_2^2}{25} + \frac{3p_1 p_2^2}{100} + \frac{p_2^3}{50} + \frac{41p_3}{25} -\frac{8p_1 p_3}{25} + \frac{3p_1^2 p_3}{50} \nonumber \\
    & \quad - \frac{21 p_2 p_3}{25} + \frac{p_1 p_2 p_3}{25} + \frac{p_2^2 p_3}{25} -\frac{39p_3^2}{50} -\frac{p_1 p_3^2}{50} -\frac{p_2 p_3^2}{25} -\frac{p_3^3}{50} + \frac{103p_4}{25} -\frac{21p_1 p_4}{100}  + \frac{3p_1^2 p_4}{100} \nonumber \\
    & \quad -\frac{13p_2 p_4}{25} + \frac{p_1 p_2 p_4}{50} + \frac{p_2^2 p_4}{50} -\frac{93p_3 p_4}{100} -\frac{p_1 p_3 p_4}{50} -\frac{p_2 p_3 p_4}{25} -\frac{3p_3^2 p_4}{100} -\frac{18p_4^2}{25} -\frac{3p_1 p_4^2}{100} -\frac{3p_2 p_4^2}{50} \nonumber \\
    & \quad -\frac{9 p_3 p_4^2}{100} -\frac{p_4^3}{25} + \frac{42p_5}{5}
\end{align}

For SU(6), 
\begin{align}
    \sum_{i<j<k}d^{abc}T_i^a T_j^b T_k^c &= \frac{5p_1}{54} - \frac{5p_1^2}{36} + \frac{5p_1^3}{108} + \frac{14p_2}{27} -\frac{7p_1 p_2}{18} + \frac{p_1^2 p_2}{9} -\frac{5p_2^2}{9} + \frac{p_1 p_2^2}{18} + \frac{p_2^3}{27} + \frac{16p_3}{9} -\frac{5p_1 p_3}{12} + \frac{p_1^2 p_3}{12} \nonumber \\
    & \quad - \frac{13 p_2 p_3}{12} + \frac{p_1 p_2 p_3}{12} + \frac{p_2^2 p_3}{12} -p_3^2 -\frac{118p_4}{27} -\frac{13p_1 p_4}{36} +\frac{p_1^2 p_4}{18} - \frac{8p_2 p_4}{9} +\frac{p_1 p_2 p_4}{18} + \frac{p_2^2 p_4}{18} \nonumber \\
    & \quad -\frac{19p_3 p_4}{12} - \frac{11 p_4^2}{9} - \frac{p_1 p_4^2}{36} -\frac{p_2 p_4^2}{18} -\frac{p_3 p_4^2}{12} -\frac{p_4^3}{27} +\frac{475p_5}{54} -\frac{2p_1 p_5}{9} +\frac{p_1^2 p_5}{36} -\frac{19p_2 p_5}{36} +\frac{p_1 p_2 p_5}{36} \nonumber \\
    & \quad  +\frac{p_2^2 p_5}{36} - \frac{11p_3 p_5}{12} -\frac{25p_4 p_5}{18} -\frac{p_1 p_4 p_5}{36} - \frac{p_2 p_4 p_5}{18} - \frac{p_3 p_4 p_5}{12} - \frac{p_4^2 p_5}{18} -\frac{35p_5^2}{36} -\frac{p_1 p_5^2}{36} - \frac{p_2 p_5^2}{18} \nonumber \\
    & \quad - \frac{p_3 p_5^2}{12} - \frac{p_4 p_5^2}{9} -\frac{5p_5^3}{108} +\frac{140 p_6}{9}
\end{align}
\end{widetext}

Note that the eigenvalue of $d$-type three-body confinement potentials for SU(4), SU(5) and SU(6) are linear in $p_4$, $p_5$ and $p_6$, respectively. 

\end{appendix}

\end{document}